\documentclass[a4paper,12pt]{article}
\pdfoutput = 1
\usepackage{jheppub}
\usepackage{amsmath,amssymb,amscd,braket,amsfonts}
\usepackage{color}
\usepackage{graphicx}
\usepackage{slashed}
\usepackage{amsthm}

\usepackage{tikz}
\usetikzlibrary{arrows.meta,calc,decorations.pathreplacing}

\newcommand{\bea}{\begin{eqnarray}}
\newcommand{\eea}{\end{eqnarray}}
\newcommand{\be}{\begin{equation}}
\newcommand{\ee}{\end{equation}}
\newcommand{\ba}{\begin{align}}
\newcommand{\ea}{\end{align}}

\usepackage[most]{tcolorbox}
\newtcolorbox{keyformula}[1][]{
  enhanced,
  breakable,
  colback=gray!4,
  colframe=black!55,
  boxrule=0.5pt,
  arc=1.5mm,
  left=2mm,
  right=2mm,
  top=1mm,
  bottom=1mm,
  fonttitle=\bfseries,
  title={#1},
  before={\par\addvspace{\abovedisplayskip}\noindent},
  after={\par\addvspace{\belowdisplayskip}\noindent\ignorespaces}
}

\newcommand{\bouncingpic}[1]{
\begin{tikzpicture}[
  >=Latex,
  line cap=round,
  line join=round,
  boundary/.style={very thick},
  ray/.style={->, very thick, red!75!black},
  raycont/.style={-, very thick, red!75!black, dotted},
  meas/.style={thin},
  tick/.style={thin},
  scale=#1
]

\begin{scope}[xshift=-7cm]

  \node[font=\bfseries] at (0,4.7) {small AdS black hole};

  \draw[boundary] (-2.5,-3.7) -- (-2.5,3.7);
  \draw[boundary] ( 2.5,-3.7) -- ( 2.5,3.7);
  \draw[boundary] (0, 3.7) ellipse (2.5 and 0.48);
  \draw[boundary] (0,-3.7) ellipse (2.5 and 0.48);

  \node[rotate=90] at (-2.95,0) {$\partial \mathrm{AdS}$};

  \fill (-0.18,-3.7) rectangle (0.18,3.7);
  \fill (0, 3.7) ellipse (0.18 and 0.08);
  \fill (0,-3.7) ellipse (0.18 and 0.08);

  \coordinate (P0) at (2.5,-2.40);
  \coordinate (P1) at (2.5, 0.00);
  \coordinate (P2) at (2.5, 2.40);

  \coordinate (B0) at (0.18,-1.20);
  \coordinate (B1) at (0.18, 1.20);
  \coordinate (MID) at (1.16, 3.00);

  \fill[red!75!black] (P0) circle (1.7pt);
  \fill[red!75!black] (P1) circle (1.7pt);
  \fill[red!75!black] (P2) circle (1.7pt);

  \draw[ray] (P0) -- (B0);
  \draw[ray] (B0) -- (P1);
  \draw[ray] (P1) -- (B1);
  \draw[ray] (B1) -- (P2);
  \draw[raycont] (P2) -- (MID);

  \node[anchor=west] at (2.62,-2.40) {$1$};
  \node[anchor=west] at (2.62, 0.00) {$\mathcal R_{\ell}$};
  \node[anchor=west] at (2.55, 2.6) {$\mathcal R_{\ell}^2$};

  \draw[meas] (4.20,-2.40) -- (4.20,2.40);
  \draw[tick] (4.07,-2.40) -- (4.33,-2.40);
  \draw[tick] (4.07, 0.00) -- (4.33, 0.00);
  \draw[tick] (4.07, 2.40) -- (4.33, 2.40);

  \node[anchor=west] at (4.40,-1.20) {$2r_*^b$};
  \node[anchor=west] at (4.40, 1.20) {$2r_*^b$};

  \node at (0,-5.4) {$\epsilon=\dfrac{R_h}{L}\ll1$};

\end{scope}

\draw[->,very thick]
  (-1.25,0) -- (1.25,0)
  node[midway,above=6pt] {$\epsilon\to0$};

\begin{scope}[xshift=2.5cm]

  \node[font=\bfseries] at (0,4.7) {Schwarzschild};

  \fill (-0.18,-3.7) rectangle (0.18,3.7);
  \fill (0, 3.7) ellipse (0.18 and 0.08);
  \fill (0,-3.7) ellipse (0.18 and 0.08);

  \coordinate (P0) at (2.5,-2.40);
  \coordinate (P1) at (2.5, 0.00);

  \coordinate (B0) at (0.18,-1.20);

  \draw[ray] (P0) -- (B0);
  \draw[ray] (B0) -- (P1);

\end{scope}
\end{tikzpicture}
}

\title{Schwarzschild scattering from CFT} 

\author{Paolo Arnaudo and}
\author{Benjamin Withers}
\affiliation{Mathematical Sciences and STAG Research Centre, University of Southampton, Highfield, Southampton SO17 1BJ, UK}
\emailAdd{p.arnaudo@soton.ac.uk}
\emailAdd{b.s.withers@soton.ac.uk}

\abstract{
We show that retarded CFT correlators on $\mathbb{R}\times \mathbb{S}^2$ in a state described by the small black hole in AdS$_4$ can be used to derive asymptotically flat Schwarzschild partial-wave S-matrix elements, including its spectrum of quasinormal modes.
This is achieved by identifying a class of operators for which the dual bulk fields behave as plane waves near the AdS boundary for sufficiently high frequencies. 
For such operators there is a simple analytic relation between CFT correlators and reflection coefficients which behaves gracefully in the flat space limit, $L\to \infty$.
These include spin-1, spin-2, and conformally coupled scalars.
We provide a simple toy model that captures how dissipative QNMs can appear for short times inside Hermitian systems, closely resembling the small black hole in AdS.
}

\begin{document}
\maketitle

\section{Introduction}

Black holes and their dynamical properties are being observed with ever-increasing precision through gravitational waves. Following a merger event, the remnant black hole rings down with characteristic complex frequencies known as quasinormal modes (QNMs), evidence of which is seen in gravitational waveforms \cite{LIGOScientific:2016lio, Carullo:2019flw, LIGOScientific:2025rid, LIGOScientific:2025wao}, for reviews see \cite{Berti:2009kk, Berti:2025hly}. These frequencies are intrinsic to the remnant black hole itself and appear in linear response observables such as the bulk-to-bulk Green's function, as well as scattering observables like partial-wave S-matrix elements on the black hole background.

In this work we show that the partial-wave S-matrix element for the Schwarzschild spacetime can be obtained from large-$N$ CFT$_3$. In particular, Schwarzschild scattering data can be extracted from CFT correlators on the Lorentzian cylinder $\mathbb{R}\times \mathbb{S}^2$ in a state corresponding to the small black hole saddle in AdS$_4$. The flat-space scattering response then appears in a particular $L\to\infty$ limit of correlators in this state, where $L$ is the radius of the $\mathbb{S}^2$, while holding the black hole size fixed and zooming in to a particular window of time, as we will describe below. Since the small black hole does not dominate the canonical ensemble our result is not directly about thermal states, however the state can instead be prepared in CFT through gravitational collapse in the bulk \cite{Bhattacharyya:2009uu}. 

By following this procedure we can recover asymptotically flat black hole scattering observables from CFT data, including the well-known Schwarzschild QNM frequencies for $s=\ell = 2$ gravitational perturbations. When combined with OPE=QNM \cite{Arnaudo:2026axe} we are also able to obtain Schwarzschild QNMs directly from OPE data for CFT$_3$.

The underlying mechanism we use is simple, and in the remainder of this introduction we provide the intuitive argument. A rigorous proof is provided in Appendix \ref{app:proof}. Consider perturbations in asymptotically AdS$_4$ spacetime governed by the following master fields obeying the Schroedinger equation
\be
\frac{\mathrm{d}^2\phi}{\mathrm{d}r_*^2}+\left[\omega^2-V(r)\right]\phi=0, \label{schroedinger}
\ee
where $r_*$ is a tortoise coordinate.
Near the boundary at $r_* = r_*^b$ for certain classes of perturbations (discussed in Section \ref{sec:operators}) one has
\be
\lim_{r_* \to r_*^b} V(r) = \frac{a}{L^2}, \label{potentialcondition}
\ee
where $L$ is the AdS radius and $a$ is a constant that is either $L$-independent or becomes $L$-independent as $L\to \infty$. Since $V(r)$ asymptotes to a constant, we can glue the bulk Schroedinger problem into a larger auxiliary problem in which we match the potential continuously at $r_* = r_*^b$ to an enlarged `lab' region $r_* > r_*^b$ with a constant potential $V_\text{lab} = V(r_*^b)$. This is possible because the potential does not diverge at the boundary for this class of perturbations and is an ordinary point of the equation.

\begin{center}
\begin{tikzpicture}[
  x=1cm,y=1cm,>=Latex,
  line cap=round,line join=round,font=\small,
  incoming/.style={draw=blue!65!black,line width=1.2pt},
  outgoing/.style={draw=red!70!black,line width=1.2pt}
]
  \filldraw[fill=blue!5,draw=black!70,line width=1pt]
    (0,0) circle (2.65);
  \node[font=\large] at (0,1.75) {AdS};
  \node[rotate=45,inner sep=3pt]
    at (-2.07,2.07) {$\partial\mathrm{AdS}$};

  \fill[black!85] (0,0) circle (0.34);
  \node[font=\footnotesize] at (0,-1.12) {black hole};

  \draw[incoming,->] (3.15,-0.70) -- (1.15,-0.70);
  \draw[incoming,samples=190,domain=3.15:7.45]
    plot (\x,{-0.70+0.115*sin(360*(\x-3.15)/0.5375)});
  \draw[incoming,->] (7.85,-0.70) -- (7.45,-0.70);

  \draw[outgoing] (1.15,-0.70)
    .. controls (0.62,-0.70) and (0.62,0.70) .. (1.15,0.70)
    -- (3.15,0.70);
  \draw[outgoing,->] (1.55,0.70) -- (2.13,0.70);
  \draw[outgoing,samples=190,domain=3.15:7.45]
    plot (\x,{0.70+0.115*sin(360*(\x-3.15)/0.5375)});
  \draw[outgoing,->] (7.45,0.70) -- (7.85,0.70);

  \draw[incoming,->] (1.15,-0.70) -- (0.31,-0.19);

  \node[font=\large] at (5.55,2.70) {Auxiliary `lab'};
  \node at (5.55,2.13) {$V_{\rm lab}=a/L^2$};
  \node[text=red!70!black] at (5.55,1.30)
    {$A\,\mathcal R_\ell\,e^{i\kappa r_*}$};
  \node[text=blue!65!black] at (5.55,-1.30)
    {$A\,e^{-i\kappa r_*}$};
  \node[font=\footnotesize,text=red!70!black,anchor=west]
    at (7.67,1.30) {out};
  \node[font=\footnotesize,text=blue!65!black,anchor=west]
    at (7.67,-1.30) {in};
  \node at (5.55,-2.40) {$\kappa^2=\omega^2-a/L^2>0$};
\end{tikzpicture}
\end{center}

We can then perform a scattering experiment from the lab, where we send plane waves with $\omega^2 > a/L^2$ incident to the AdS region and measure the reflectivity, $\mathcal{R}_\ell$. The matching of solutions across $r_*^b$ defines a near-boundary basis of solutions in AdS, $u_\pm(r_*)$, which are defined through the following boundary conditions,
\be
u_\pm(r_*^b)=e^{\pm i\kappa r_*^b}, \qquad \partial_{r_*} u_\pm(r_*^b)= \pm i\kappa e^{\pm i\kappa r_*^b}, \label{bdybasisdef}
\ee
where $\kappa = \sqrt{\omega^2- a/L^2}$. Consequently these define a basis of solutions in AdS, and we can write
\be
\phi = A\left(u_- + \mathcal{R}_\ell u_+\right). \label{planewaves}
\ee
Here $\mathcal{R}_\ell$ is the reflection coefficient for the lab scattering problem. The key point with this auxiliary construction is that as $L\to \infty$ the lab scattering problem reduces to the asymptotically flat black hole scattering problem, and $\mathcal{R}_\ell$ becomes the reflection coefficient for the black hole as a partial-wave S-matrix element.\footnote{For a study on the singularity structure in the complex frequency plane of the scattering S-matrix see \cite{Correia:2025enx}.}

Using the representation \eqref{planewaves} and their near-boundary behaviour \eqref{bdybasisdef} we can directly compute the CFT correlator, $\widetilde{G}(\omega)$, which gives a relation between it and $\mathcal{R}_\ell$, up to normalisation and contact terms,\footnote{For some perturbation types, this relation will need to be adjusted, since the map between master fields and fundamental fields changes the relation. This will be discussed further in section \ref{sec:operators}.}
\be
L^{2\Delta - d}\widetilde{G}(\omega) \equiv L\frac{\partial_{r_*}\phi}{\phi}\bigg|_{r_* =r_*^b} = i \kappa L \frac{\mathcal{R}_\ell e^{2i \kappa r_*^b}-1}{\mathcal{R}_\ell e^{2i \kappa r_*^b}+1}. \label{bcrelation}
\ee
While the lab experiment defines a single scattering event from the black hole, with radiation escaping to infinity in the lab, the CFT correlator \eqref{bcrelation} defines the response properties in the AdS box subject to Dirichlet boundary conditions, where radiation cannot escape. When $\mathcal{R}_\ell$ is small, this relation can be expanded as a geometric series
\be
L^{2\Delta - d}\widetilde{G}(\omega) = -i\kappa L + 2 i \kappa L \mathcal{R}_\ell e^{2i \kappa r_*^b} + \ldots. \label{expanded}
\ee
The first term in this expression corresponds to prompt response. Higher terms correspond to reflections from the interior and subsequent sequence of reflections between the interior and the boundary. Thus, in cases where \eqref{potentialcondition} holds, the CFT correlator can be viewed as the resummation of successive scattering events between the black hole and the AdS boundary, as depicted below.

\begin{center}
\bouncingpic{0.7}
\end{center}

\begin{figure}[h!]
\centering
\includegraphics[width=\linewidth]{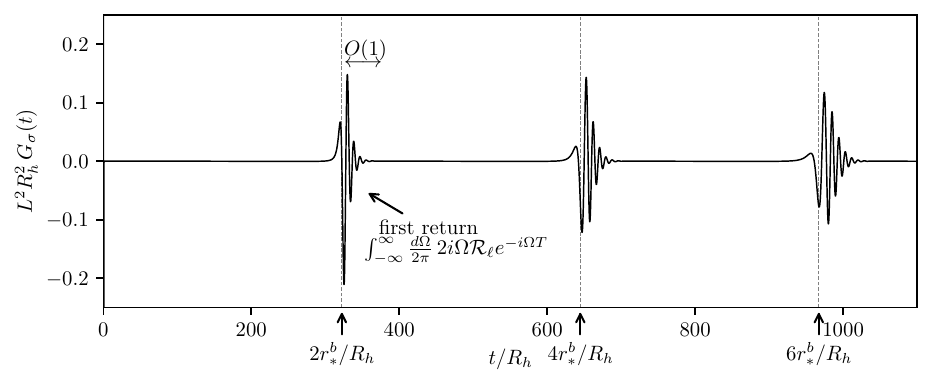}
\caption{An approximation $G_\sigma(t)$ of the time domain CFT correlator $G(t)$ obtained by solving \eqref{schroedinger} with Gaussian boundary source of width $\sigma$ centred at $t=0$. Here $\epsilon = R_h/L = 10^{-2}$ and $\sigma = 0.5$. The wavepacket bounces back and forth between the bulk and the boundary, corresponding to pulses appearing in the correlator at return times $2n r_*^b$. A scaling limit of the first return yields asymptotically flat Schwarzschild scattering in accordance with \eqref{keyresult}. \label{fig:Gt}} 
\end{figure}

We now wish to extract a single scattering event in the case of the 3+1 dimensional black hole spacetime in the flat space limit $L\to \infty$, instead of bouncing back and forth in the finite-$L$ AdS box. First, since $\mathcal{R}_\ell$ as defined through \eqref{planewaves} depends on the definition of $r_*$, we must fix its ambiguous integration constant. We do this such that in the $L\to \infty$ limit $\mathcal{R}_\ell$ becomes the canonical definition for the reflection coefficient in the Schwarzschild geometry.\footnote{Specifically, through comparison to the asymptotically flat tortoise coordinate $r_* = r + 2M \log\left(\frac{r}{2M}-1\right)$.}
This is achieved by setting
\be
r_*^b = \frac{\pi L}{2} + R_h \log\left(\frac{L}{R_h}\right) - \frac{R_h}{2}.
\ee
The first term is the time it takes for a signal to reach the centre of empty AdS from the boundary, while the second term is the associated Shapiro time delay caused by the presence of the black hole. Thus in the large $L$ limit, $2r_*^b$ becomes the reference time for the first return of a signal from the bulk. The third term simply ensures the correct phase shift for comparison with the Schwarzschild result. Next, we need to extract a single term from \eqref{expanded}. One way to achieve this is in the time domain,
\begin{equation}
    G(t) = \int_{-\infty}^\infty \frac{d\omega}{2\pi}\widetilde{G}(\omega) e^{-i \omega t}.
\end{equation}
For intuition, we present the structure of the time domain correlator for axial gravitational perturbations at large but finite $L$ in Figure \ref{fig:Gt}, obtained through numerical integration.

At large $L$ \eqref{expanded} becomes
\be
L^{2}G(t)= \delta'(t)
+\int_{-\infty}^{\infty}\frac{d\omega}{2\pi}\,
2i\omega\mathcal{R}_\ell e^{-i\omega(t-2r_*^b)}+\ldots.
\ee
The key point is to recognise that the contribution for the first return is localised near a time of $t = 2r_*^b$, and its nontrivial time dependence lasts for a time of order $R_h$ where $R_h$ is the small black hole size. Accordingly, taking the limit $L \to \infty$ with the following dimensionless quantities held fixed,
\be
T = \frac{t-2r_*^b}{R_h}, \qquad \Omega=\omega R_h
\ee
isolates the first return contribution, 
\begin{keyformula}
\begin{equation}
\lim_{R_h/L\to 0} L^2R_h^2G(2r_*^b+R_hT) = \int_{-\infty}^{\infty}\frac{d\Omega}{2\pi}\, 2i\Omega\mathcal{R}_\ell e^{-i\Omega T}. \label{keyresult}
\end{equation}
\end{keyformula}
This is a key result of this work: on the left hand side we have a limit of a CFT$_3$ correlator, while on the right hand side we have a Fourier transform of a asymptotically flat Schwarzschild reflection coefficient. 

We note that previous work \cite{Cardoso:2015fga} has examined the bulk Green's function for small black holes in AdS, showing that probing the black hole from a finite location in the bulk for short enough times is unaffected by the AdS asymptotics. Instead, our work concerns the appearance of Schwarzschild ringdown in the CFT correlator which is an asymptotic observable, defined at the AdS conformal boundary. 

There is also a long established literature connecting AdS correlators with S-matrix elements \cite{Polchinski:1999ry, Susskind:1998vk, Giddings:1999jq, Gary:2009ae, Gary:2009mi, Fitzpatrick:2010zm, Penedones:2010ue, Fitzpatrick:2011ia, Hijano:2019qmi, Marotta:2024sce, Berenstein:2025tts}. These works predominantly focus on vacuum correlators and propagation through empty AdS, rather than in perturbation theory around an already-prepared black hole state as we do here. Our work shares similarities with recent constructions of duals to worldline effective theories in the bulk \cite{Correia:2026fnr}.

The rest of the paper is arranged as follows. In Section \ref{sec:operators} we discuss the types of perturbation for which \eqref{potentialcondition} holds. In Section \ref{sec:extraction} we then demonstrate numerically that in the case of $s=\ell = 2$ axial perturbations we recover the correct Schwarzschild QNM spectrum from the CFT correlator. We discuss obtaining Schwarzschild from OPE data in Section \ref{sec:OPEQNM}. In section \ref{sec:toymodel} we show how much of the dynamics we observe, namely the appearance of damped QNMs in CFT response functions near the return time, can be reproduced by a simple toy model consisting of a piece of string coupled to an oscillator at one end. We finish with a discussion in section \ref{sec:discussion}.

\section{Operators \label{sec:operators}}
A key property of the above result was the potential asymptoting to a constant at the AdS boundary, \eqref{potentialcondition}. For which types of bulk fields, and therefore CFT primaries, does this happen?

We can obtain a unified treatment for scalars, electromagnetic and gravitational perturbations following Ishibashi-Wald \cite{Ishibashi:2004wx} in the case of empty AdS$_{d+1}$,
\begin{equation}
    ds^2 = \frac{L^2}{\sin^2x}\left(-dt^2 + dx^2 + \cos^2x d\Omega_{d-1}^2\right)
\end{equation}
for which the master field equation takes the form
\be
\frac{\mathrm{d}^2\Phi}{\mathrm{d}x^2} + \left[\omega^2 - V(x)\right]\Phi = 0,\qquad V(x) = \frac{\nu^2-1/4}{\sin^2x} + \frac{\sigma^2-1/4}{\cos^2x},
\ee
where $\sigma = \ell + \frac{d-2}{2}$ and $\nu^2$ depends on the species. The condition of interest, therefore is $\nu^2 = 1/4$. For scalars, $(\Box - m^2)\phi = 0$, one has
\begin{equation}
    \nu^2 - \frac{1}{4} = \frac{(d-1)(d+1)}{4} + m^2 L^2
\end{equation}
which vanishes only in the case of a conformally coupled scalar in any dimension $d$. Note that this corresponds to operator dimensions $\Delta_\pm = \frac{d}{2}\pm \frac{1}{2}$ and either quantisation choice is permissible. For electromagnetic peturbations,  
\begin{eqnarray}
    \nu^2 - \frac{1}{4} &=& \frac{(d-1)(d-3)}{4}\qquad \text{(vector-type)}\\
    \nu^2 - \frac{1}{4} &=& \frac{(d-3)(d-5)}{4}\qquad \text{(scalar-type)}
\end{eqnarray}
so vector-type in $d=3$, and scalar-type in $d=3$ and $d=5$ satisfy the condition. Finally, for gravitational peturbations we have
\begin{eqnarray}
    \nu^2 - \frac{1}{4} &=& \frac{(d-1)(d+1)}{4}\qquad \text{(tensor-type)}\\
    \nu^2 - \frac{1}{4} &=& \frac{(d-1)(d-3)}{4}\qquad \text{(vector-type)}\\
    \nu^2 - \frac{1}{4} &=& \frac{(d-3)(d-5)}{4}\qquad \text{(scalar-type)}
\end{eqnarray}
leading to vector-type with $d=3$ (axial/odd) and scalar-type with $d=3$ (polar/even) and $d=5$ only.

Moving now to the case of interest of the AdS$_4$ black hole spacetime, the corresponding potentials for our master equation \eqref{schroedinger} are as follows
\begin{eqnarray}
    V^{s=0}_{\rm conf.}(r) &=& f(r)\left[\frac{\ell(\ell+1)}{r^2} +\frac{2M}{r^3}\right],\\
    V^{s=1}(r) &=& f(r)\frac{\ell(\ell+1)}{r^2},\\
    V^{s=2}_{\rm odd}(r) &=& f(r)\left[\frac{\ell(\ell+1)}{r^2} - \frac{6M}{r^3}\right],\\
    V^{s=2}_{\rm even}(r) &=& \frac{2f(r)}{r^3(\lambda r+3M)^2} \left[ \lambda^2(\lambda+1)r^3 +3\lambda^2Mr^2 +9\lambda M^2r +9M^3 +\frac{9M^2r^3}{L^2} \right], \nonumber \\
\end{eqnarray}
where in the last line, $\lambda = \frac{1}{2}(\ell-1)(\ell+2)$.

Note that in most cases above, the map between asymptotic data in the master field and the CFT correlator is given as in \eqref{bcrelation}. However for certain types of perturbation this relation is adjusted due to the way the master fields relate to the fundamental fields. The first case is the conformally coupled scalar with alternative choice of boundary conditions, \cite{Klebanov:1999tb}, and in this case the relation is inverted. The second case is electromagnetic scalar-type perturbations in $d=3$ \cite{Ishibashi:2004wx, Herzog:2007ij}, again, inverting the relation. Finally, the gravitational scalar-type (polar/even) perturbations in $d=3$ require additional care \cite{Bakas:2008gz, Bakas:2009pbm}. Performing these adjustments does not qualitatively change the nature of the argument. In this work we focus primarily on the gravitational vector-type (axial/odd) perturbations for which no adjustment is required.

\section{Schwarzschild QNMs\label{sec:extraction}}
In this section we illustrate the main result \eqref{keyresult} by extracting the QNM frequencies of the asymptotically flat Schwarzschild spacetime for $s=2$, $\ell = 2$ axial perturbations directly from the CFT$_3$ correlator. These frequencies appear as poles in $\mathcal{R}_\ell$, and thus will be evident in the time domain signal close to the return time.

We first construct the CFT correlator numerically in time domain, by solving the PDE corresponding to \eqref{schroedinger} with $\omega^2$ replaced by $-\partial_t^2$. In the time domain the correlator results from imposing the boundary condition, 
\be
\phi(t, r= r^b) = R_h\delta(t),
\ee
together with ingoing behaviour at the horizon, and then reading off the Neumann data,
\be
R_h^2L^2G(t) = R_h\partial_{r_*}\phi \big|_{r_* = r_*^b}.
\ee
In practice we approximate the delta function by a normalised Gaussian of width $\sigma$ for the purposes of numerical evolution, and denote the corresponding approximated correlator by $G_\sigma(t)$. The evolution is performed using second order finite differences in space and explicit leapfrog stepping in time, and importantly for the results of this section, using high-precision arithmetic. A representative result from evolution is shown in Figure \ref{fig:Gt} at $\epsilon = 10^{-2}$.

The resulting profile for $G_\sigma(t)$ shown in Figure \ref{fig:Gt} displays the ringdown profile following a time of $2r_*^b$. We fit this behaviour to damped exponentials using the Prony method. This method is a particularly robust way of spectrally decomposing a time domain signal, and was utilised for this purpose in \cite{Arnaudo:2026axe}. We choose a fitting window $t \in [t_0, t_{N-1}]$ containing the $N$ discretised time samples at $t_n$, and approximate the signal as a finite sum of complex exponentials over this window:
\be
G_\sigma(t_n) \simeq \sum_{j=1}^N A_j e^{-i \omega_j(t_n - t_0)}.
\ee
From this two Hankel matrices of size $(N-p)\times p$ where $p=\left \lfloor{N/2}\right \rfloor $ are constructed as follows: $(H_0)_{ij} = G_\sigma(t_{i+j})$ and $(H_1)_{ij} = G_\sigma(t_{i+j+1})$. From this we construct the following SVD, $H_0 = U \Sigma V^\dagger$, and retain the leading $r$ singular components from which we form $M_r = U_r^\dagger H_1 V_r \Sigma_r^{-1}$. The spectrum is then given by $\omega_j = \frac{i}{\delta t}\log z_j$, where $z_j$ are the eigenvalues of $M_r$. 

\begin{figure}[h!]
\centering
\includegraphics[width=0.7\linewidth]{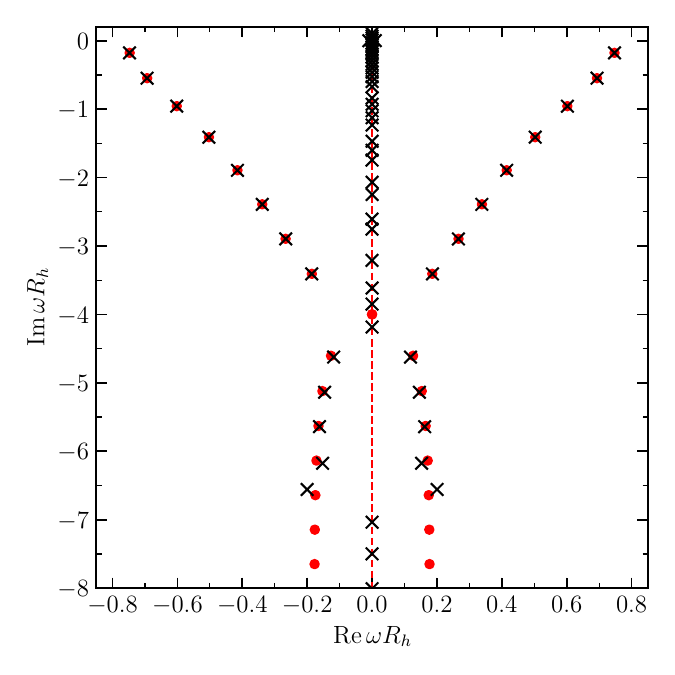}
\caption{Spectral decomposition of a CFT correlator after a time corresponding to the first return for a small black hole in AdS$_4$ showing agreement with the asymptotically flat Schwarzschild black hole QNM spectrum. Black crosses are complex frequencies extracted from Prony fits to $G_\sigma(t)$ as in Figure \ref{fig:Gt}, while red dots are the Schwarzschild QNMs taken from \cite{Berti:2005ys, Berti:2009kk}. Two Prony windows were used: $T \in [10,30]$ and $T \in [15,35]$ fitting 250 modes, and keeping those modes whose magnitudes differ by less than $0.1$. \label{fig:pdeprony}} 
\end{figure}

The results for the Prony fitting procedure are shown in Figure \ref{fig:pdeprony} for a time window restricted to the ringdown behaviour following the first return time. For this window, there is a clear agreement with the asymptotically-flat Schwarzschild black hole QNMs. In addition we see that the Prony method yields a number of modes along the imaginary frequency axis. These represent the contribution of the branch cut contribution in Schwarzschild that results from trying to approximate a power-law tail by a sum of exponentials (see e.g. \cite{Besson:2024adi, Arnaudo:2025kit}), and its precise resolution is sensitive to the Prony window and parameters. These findings are consistent with our key result \eqref{keyresult} which applies in the $L\to \infty$ limit, zoomed in on this particular time window. 

Of course, the correlator we have computed here numerically is not at infinite $L$. This means that the `true' QNMs of the present finite-$L$ system are those of a small black hole in AdS, corresponding almost normal modes, with small imaginary part \cite{Aminov:2023jve}. This spectral behaviour is not evident in the small-window fit that we have presented in this section, and applies instead at large times. This will be discussed in Section \ref{sec:OPEQNM}.

\section{OPE = QNM \label{sec:OPEQNM}}
CFTs admit operator product expansions (OPE) when operators are brought close together. The microscopic data that appears in these expansions are often used as the defining set of data for the CFT itself. Indeed, the OPE is a crucial ingredient in the conformal bootstrap programme, allowing precise determination of operator dimensions or identifying new candidate CFTs \cite{Poland:2018epd}. 

While the small black hole state we study does not dominate the canonical ensemble, the structure of the OPE in this case still takes on the form of a thermal OPE with inverse temperature scale $\beta$, as the symmetries of stationarity and rotational invariance are still the same. In such systems there are two relevant expansions: the OPE which is valid on timescales small with respect to $\beta$, and the QNMs of the system which should be regarded as emergent collective phenomena. As explored in \cite{Arnaudo:2026axe}, these two representations admit an overlapping region of convergence, leading to the slogan `OPE = QNM'. See also \cite{Barrat:2026jfg} for closely related work.

A natural question therefore is whether the asymptotically flat Schwarzschild reflection coefficient and associated QNMs can be recovered purely from the thermal OPE data of the CFT. Through OPE = QNM, a closely related question is whether the QNMs of the Schwarzschild-AdS$_4$ spacetime can reconstruct the QNMs of the asymptotically flat Schwarzschild spacetime.
In this section we answer both questions in the affirmative. 

The Schwarzschild-AdS QNM frequencies and residues were computed directly from the Regge--Wheeler equation. After analytically continuing the ingoing solution at the horizon to the boundary, the QNM frequencies are determined as the poles of the boundary logarithmic derivative $L^2\widetilde G(\omega)=\partial_{r_*}\phi_{\mathrm{in}}(r_*^b,\omega)/\phi_{\mathrm{in}}(r_*^b,\omega)$.
For a simple pole $\omega_n$, the corresponding residue is
\begin{equation}
r_n=\frac{\partial_{r_*}\phi_{\mathrm{in}}(r_*^b,\omega_n)}{\partial_\omega\phi_{\mathrm{in}}(r_*^b,\omega_n)}.
\end{equation}
The OPE data were recursively generated independently by expanding the radial logarithmic derivative at large frequency, and then converting them to a short-time expansion via a term-by-term inverse Fourier transform. The leading term, $-\mathrm{i}\omega$ produces the contribution $\delta'(t)$, while the remaining coefficients determine the regular time series.

\begin{figure}[h!]
\centering
\includegraphics[width=\linewidth]{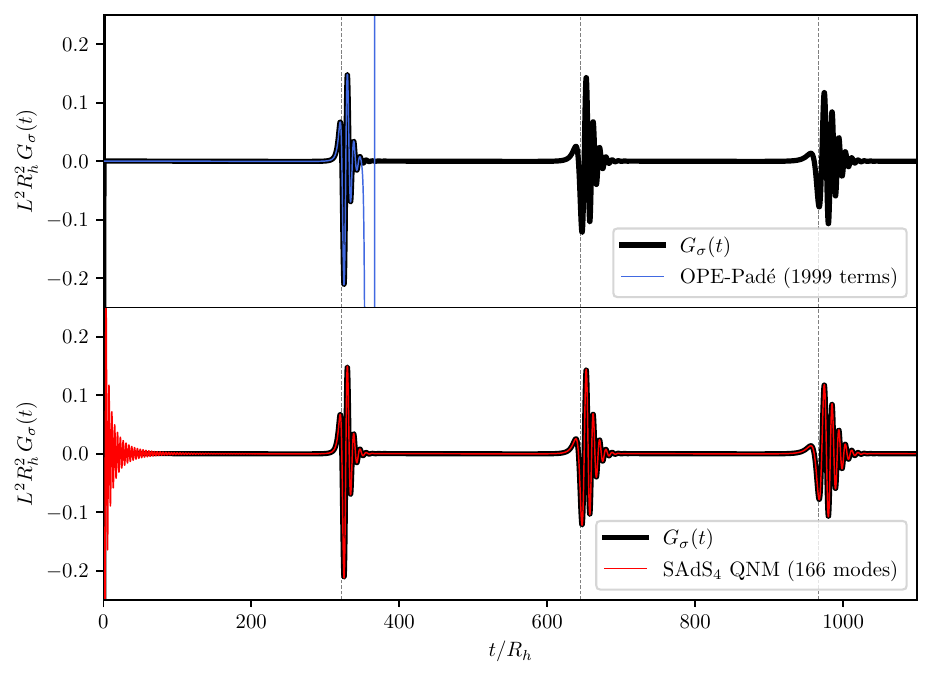}
\caption{Reproducing asymptotically flat Schwarzschild scattering from different sets of CFT data. \textbf{Upper panel}: From OPE data, analytically continued using a diagonal Pad\'e approximant in the variable \eqref{wvar}, from 1999 time-domain OPE coefficients (blue). 
\textbf{Lower panel}: From the QNM expansion of the CFT correlator. These are the almost-normal modes of the small black hole in AdS. We show the QNM residue sum for 83 positive-real-frequency modes and their $\omega \to -\omega^*$ partners, giving 166 modes in total (red).
Vertical dashed lines mark integer multiples of the return time $2r_*^b$, and the black curves are the CFT correlator constructed numerically, $G_\sigma(t)$, for comparison.
\label{fig:OPEQNM}
}
\end{figure}

\begin{figure}[t]
\centering
\makebox[\linewidth][c]{%
\begin{tikzpicture}[x=1cm,y=1cm,font=\small, scale=0.8]

\begin{scope}
\draw[->] (-4.2,0) -- (4.25,0);
\node[above left] at (4.25,0.18) {$\mathrm{Re}\,t$};
\draw[->] (0,-2.55) -- (0,2.6);
\node[above right] at (0,2.6) {$\mathrm{Im}\,t$};

\draw[gray!60,densely dotted] (1.2,-2.55) -- (1.2,2.15);
\node[below] at (1.2,-2.15) {$2r_*^b$};

\draw (-0.07,0.65) -- (0.07,0.65);
\draw (-0.07,-0.65) -- (0.07,-0.65);
\node[left,font=\footnotesize] at (-0.15,0.65) {$\beta/2$};
\node[left,font=\footnotesize] at (-0.15,-0.65) {$-\beta/2$};

\foreach \n in {1,2,3}{
    \fill[red!75!black] ({1.2*\n},{0.65*\n}) circle (2.3pt);
    \fill[red!75!black] ({1.2*\n},{-0.65*\n}) circle (2.3pt);
    \fill[red!75!black] ({-1.2*\n},{0.65*\n}) circle (2.3pt);
    \fill[red!75!black] ({-1.2*\n},{-0.65*\n}) circle (2.3pt);
}

\draw[red!75!black,thick] (1.2,0.65) circle (5pt);
\draw[red!75!black,thick] (1.2,-0.65) circle (5pt);

\node[red!75!black,rotate=28] at (4,2.17) {$\cdots$};
\node[red!75!black,rotate=-28] at (4,-2.17) {$\cdots$};
\node[red!75!black,rotate=-28] at (-4,2.17) {$\cdots$};
\node[red!75!black,rotate=28] at (-4,-2.17) {$\cdots$};

\node at (0,-3.6)
    {$t=n\left(2r_*^b\pm\mathrm{i}\beta/2\right)$};
\end{scope}

\begin{scope}[shift={(9.3,0)}]
\draw[->] (-4.2,0) -- (4.25,0);
\node[above left] at (4.25,0.18) {$\mathrm{Re}\,t$};
\draw[->] (0,-2.55) -- (0,2.6);
\node[above right] at (0,2.6) {$\mathrm{Im}\,t$};

\foreach \n in {-3,-2,-1,1,2,3}{
    \fill[black] ({1.2*\n},0) circle (2.3pt);
}

\node[below] at (1.2,-0.18) {$\pi L$};
\node[below] at (2.4,-0.18) {$2\pi L$};
\node[below] at (3.6,-0.18) {$3\pi L$};
\node[below] at (-1.2,-0.18) {$-\pi L$};
\node[below] at (-2.4,-0.18) {$-2\pi L$};
\node[below] at (-3.6,-0.18) {$-3\pi L$};

\node[fill=white,inner sep=1pt] at (3.95,0) {$\cdots$};
\node[fill=white,inner sep=1pt] at (-3.95,0) {$\cdots$};

\node at (0,-3.6) {$t=n\pi L$};
\end{scope}

\end{tikzpicture}%
}
\caption{Schematic complex-time singularity structure. \textbf{Left panel}: the bouncing singularities $t=n(2r_*^b\pm\mathrm{i}\beta/2)$ of a small Schwarzschild-AdS$_4$ black hole. The circled points are the first pair, $t=2r_*^b\pm\mathrm{i}\beta/2$, associated with the first return. Their common real part is the return time $2r_*^b$, around which we resolve the response by taking $t=2r_*^b+R_hT$ and holding $T$ fixed as $R_h/L\to0$. \textbf{Right panel}: the real-time return singularities of pure global AdS$_4$, at $t=n\pi L$. \label{bouncingsingAdS}}
\end{figure}
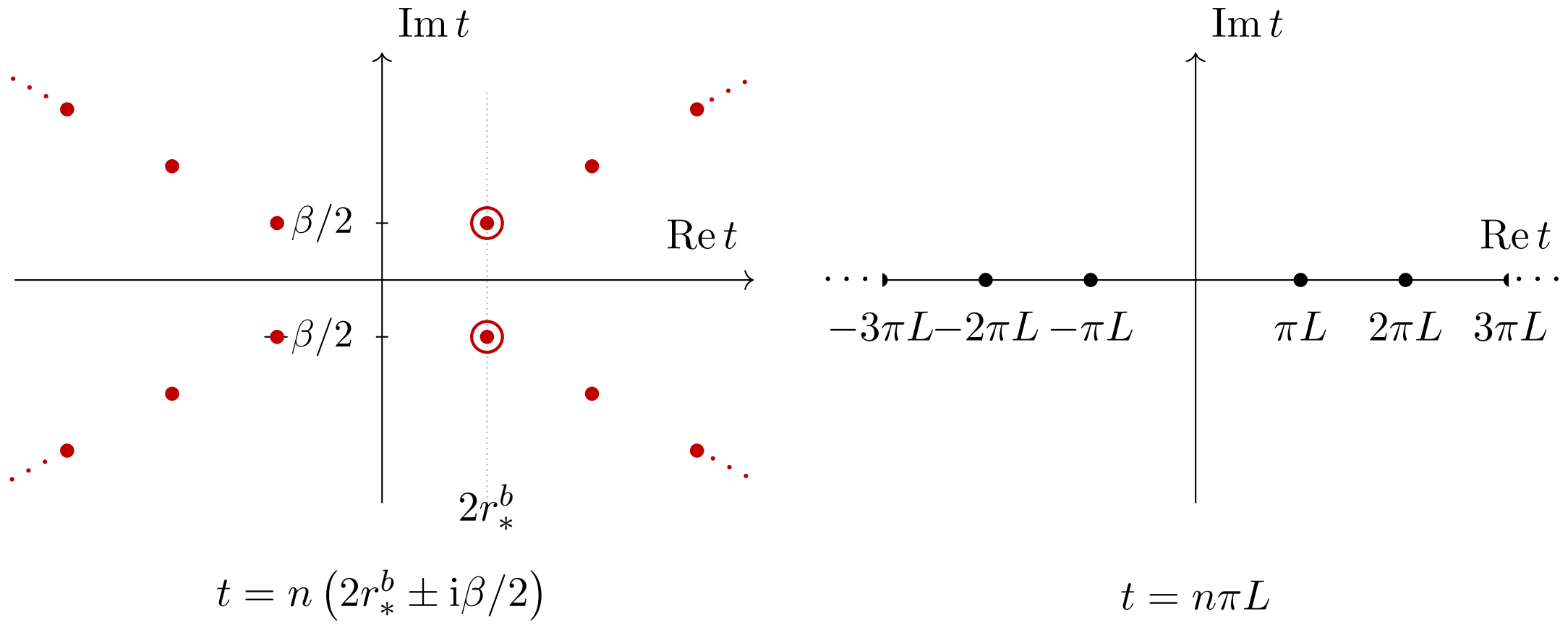

The OPE expansion is limited by complex-time bouncing singularities of $G(t)$ at $t=2r_*^b\pm\mathrm{i}\beta/2$ which approach $t = 2 r_*^b$ as $\epsilon \to 0$, as shown in Figure \ref{bouncingsingAdS}.
These are the real-time return singularities at integer multiples of $\pi L$ for the empty AdS geometry. To continue beyond these singularities we construct Padé approximants in a conformal variable adapted to the bouncing singularities:
\begin{equation}
w(t)=\frac{t}{\sqrt{(2r_*^b)^2+\beta^2/4}+\sqrt{(t-2r_*^b)^2+\beta^2/4}}.\label{wvar}
\end{equation}
This map sends the two specified complex points to the unit circle, while $t=0$ maps to $w=0$ and the positive real-time axis lies inside $0\le w<1$. At the same OPE order, we find that Padé approximants constructed in $w$ reproduce the first-return response more accurately than those constructed directly in $t$. The result is shown in the upper panel of Figure \ref{fig:OPEQNM}.

The Schwarzschild-AdS QNM sum, instead, reproduces the successive returns for all $t>0$, in accordance with the QNM sum being convergent everywhere at real times \cite{Arnaudo:2025uos}. The result is shown in the lower panel of Figure \ref{fig:OPEQNM}. The rapid oscillations near $t=0$ reflect the limited ability of the truncated QNM sum to resolve the sharply localised prompt response.

\section{A piece of string as a toy model}\label{sec:toymodel}
Many aspects of the small black hole problem can be understood through a toy model in which all quantities can be computed in closed form. The model is to take a piece of string of length $L$ and couple one end to a simple harmonic oscillator. Sending in a wave from the other end of the string excites the harmonic oscillator, which then radiates its energy back into the string. This leads to an observed short-time response after the first return in which there is damped QNM decay, representing the decay of the oscillator. However, the system as designed is Hermitian: the energy is conserved, as it is just exchanged between radiation on the string and the oscillator. Thus while the short time behaviour after a return is governed by QNMs, the long-time behaviour is governed by normal modes. This is analogous to the small black hole in AdS case where short time response is governed by asymptotically flat Schwarzschild, while the long time response is governed by the nearly-normal modes of AdS.

The model is described by the following action,
\be
S = \frac{1}{2}\int dt \int_0^L dx \left((\partial_t\phi)^2 - (\partial_x\phi)^2\right) + \frac{m}{2}\int dt \left(\left(\partial_t\phi_0\right)^2 - \omega_0^2 \phi_0^2\right)
\ee
where $\phi_0(t) = \phi(t,0)$. The equations of motion are the wave equation $\partial_t^2 \phi = \partial_x^2\phi$ subject to a boundary condition at $x=0$, $\partial_x \phi|_{x=0} = m(\partial_t^2\phi_0 + \omega_0^2 \phi_0)$. This replaces the ingoing boundary condition in the black hole case. The general solution in frequency space after imposing this boundary condition is
\be
\widetilde{\phi} = A \left( \cos(\omega x) + m(\omega_0^2 - \omega^2)\frac{\sin(\omega x)}{\omega}\right). \label{toyingoing}
\ee

The analogue of the CFT correlator in frequency space is the ratio of Neumann to Dirichlet data at $x=L$, and is given by
\be
\widetilde{G}(\omega) = L\frac{\partial_x \widetilde{\phi}}{\widetilde{\phi}}\bigg|_{x=L} = \omega L\frac{-\omega \sin(\omega L) + m(\omega_0^2 - \omega^2)\cos(\omega L)}{\omega \cos(\omega L) + m(\omega_0^2 - \omega^2)\sin(\omega L)}.
\ee
This admits only poles at real frequencies which are all simple, with exactly one lying in each interval $\frac{n\pi}{L} < \omega_n < \frac{(n+1)\pi}{L}$ with $n = 0,1, \ldots$, together with their partner $-\omega_n$. At large overtone number these modes asymptote to $\omega_n = \frac{n\pi}{L} + \frac{1}{m n \pi} + O(n^{-3})$.
The reflection coefficient is given by wave scattering from the boundary, read off from the coefficients of \eqref{toyingoing} in exponential form,
\be
\mathcal{R} = -1 + \frac{2i\omega}{i\omega + m(\omega^2 - \omega_0^2)}.
\ee
This exhibits a pair of damped QNMs, at frequencies $\omega = -\frac{i}{2m} \pm \frac{\sqrt{4 m^2 \omega_0^2 - 1}}{2m}$.
The relation between the two quantities is just as in the black hole case, \eqref{bcrelation}
\be
\widetilde{G}(\omega)  = i \omega L\frac{\mathcal{R} e^{2i L\omega} -1}{\mathcal{R} e^{2i L\omega} +1},\label{toybcrelation}
\ee
where now $2L$ corresponds to the return time. The same arguments as the black hole case now apply. Expanding \eqref{toybcrelation} in a geometric series isolates each return contribution in the inverse Fourier transform to time domain. The first few terms in this expansion correspond to 
\bea
G(t) =&& L\delta'(t)\\
&+&2L\delta'(t-2L)-8\gamma L\delta(t-2L)
+L\theta(t-2L)f_1(t-2L)\\
&+&2L\delta'(t-4L)-16\gamma L\delta(t-4L)
+L\theta(t-4L)f_2(t-4L)\\
&+& \ldots
\eea
where $\gamma=1/(2m)$, $\nu=\sqrt{\omega_0^2-\gamma^2}$, and
\bea
f_1(\tau) &=& 8\gamma e^{-\gamma\tau} \left[ 2\gamma\cos(\nu\tau) +\frac{\nu^2-\gamma^2}{\nu}\sin(\nu\tau) \right],\\
f_2(\tau) &=& 16\gamma e^{-\gamma\tau} \Bigg[ \left( 4\gamma-\gamma^2 \left(3-\frac{\gamma^2}{\nu^2}\right)\tau \right)\cos(\nu\tau) \nonumber\\
&&\qquad+ \left( \nu-\frac{4\gamma^2}{\nu}-\frac{\gamma^4}{\nu^3} +\frac{\gamma(3\gamma^2-\nu^2)}{\nu}\tau \right)\sin(\nu\tau) \Bigg].
\eea
Thus through $f_{1}(\tau)$ and $f_{2}(\tau)$ one observes ringdown after the first and second returns due to the damped `black hole' QNMs, despite the full time dependence being described by a sum of normal modes.

\section{Discussion\label{sec:discussion}}
Our results identify a precise part of CFT correlators that retains the scattering dynamics of asymptotically flat Schwarzschild black holes. The essential operation is to isolate the first return and resolve its time dependence on the horizon scale, keeping $T=(t-2r_*^b)/R_h$ fixed as $R_h/L\to0$. Subtracting $2r_*^b$ removes the long propagation through the exterior AdS region, while the remaining time dependence resolves scattering due to the black hole in AdS. In the axial channel of the stress-tensor correlators considered here, the resulting response is determined by the flat reflection coefficient, whose resonance poles give the Schwarzschild QNM frequencies.

In Section \ref{sec:operators}, we described the class of operators to which our procedure applies. Nevertheless, the mechanism can be generalised to a larger class of cases. A potential that diverges at the AdS boundary prevents the direct plane-wave construction we presented, but does not by itself obstruct the AdS-to-flat mechanism. The boundary propagation must then be treated through the full connection problem. To make this distinction explicit, suppose that the local solution ingoing at the black hole horizon has been connected to two exact solutions in some overlap region $R_h\ll r\ll L$, chosen and normalised so that they approach the outgoing and incoming flat scattering solutions in the flat limit. Let us write
\begin{equation}
\phi_{\mathrm{in}}=C_{\mathrm{in},\mathrm{up}}\phi_{\mathrm{up}}+C_{\mathrm{in},\mathrm{down}}\phi_{\mathrm{down}}.
\end{equation}
Denoting the source and response coefficients of each basis solution by $C_{\mathrm{up},\mathrm{NN}}$, $C_{\mathrm{up},N}$ and $C_{\mathrm{down},\mathrm{NN}}$, $C_{\mathrm{down},N}$, the frequency-space correlator, can still be written in the fractional-linear form
\begin{equation}\label{eq:correlator-linear-fractional}
\widetilde G(\omega)=\frac{C_{\mathrm{in},\mathrm{up}}C_{\mathrm{up},N}+C_{\mathrm{in},\mathrm{down}}C_{\mathrm{down},N}}{C_{\mathrm{in},\mathrm{up}}C_{\mathrm{up},\mathrm{NN}}+C_{\mathrm{in},\mathrm{down}}C_{\mathrm{down},\mathrm{NN}}},
\end{equation}
where, unlike \eqref{bcrelation}, its coefficients contain the nontrivial propagation between the overlap region and the AdS boundary. The same connection formula also separates contributions according to the number of reflections from the inner region. Writing $\mathcal R_\ell=C_{\mathrm{in},\mathrm{up}}/C_{\mathrm{in},\mathrm{down}}$, expansion at fixed frequency gives
\begin{equation}\label{eq:Gtildeexpansion}
\widetilde G(\omega)=\frac{C_{\mathrm{down},N}}{C_{\mathrm{down},\mathrm{NN}}}+\mathcal R_\ell\,\frac{C_{\mathrm{up},N}C_{\mathrm{down},\mathrm{NN}}-C_{\mathrm{down},N}C_{\mathrm{up},\mathrm{NN}}}{C_{\mathrm{down},\mathrm{NN}}^2}+\mathcal O(\mathcal R_\ell^2).
\end{equation}
Here $C_{\mathrm{down},\mathrm{NN}}\ne0$, and the underlying geometric series converges when
\begin{equation}
\left|\mathcal R_\ell\frac{C_{\mathrm{up},\mathrm{NN}}}{C_{\mathrm{down},\mathrm{NN}}}\right|<1.
\end{equation}
The linear term in \eqref{eq:Gtildeexpansion} describes a single reflection from the inner region, including propagation to and from the AdS boundary, with the prefactor fixed by the outer connection problem. Therefore, even though the full correlator depends on additional non-trivial connection data, the reflection coefficient can still be identified in its analytic expression. This suggests a similar structure\footnote{This is also suggested by the approach in \cite{Marotta:2024sce} which is valid also for spinning massive amplitudes.} to the one we studied in this work, however the precise extraction of the corresponding flat data presently remains obscured by the additional data.

There is an alternative interpretation of this connection procedure.\footnote{We thank Cristoforo Iossa for suggesting this viewpoint.} The connection procedure can be organised using a local Frobenius basis at either complex zero of $f(r)$, which approaches $r=\pm iL$ as $R_h/L\to0$. The ingoing solution at the horizon can first be connected to this basis and then analytically continued to the AdS boundary. For the perturbations studied here, the boundary is an ordinary point of the radial equation, so the continued solutions and their derivatives can be evaluated there directly. For other perturbations, for which the boundary is a regular singular point, an additional connection to the non-normalisable and normalisable boundary solutions is required, giving the coefficients entering \eqref{eq:correlator-linear-fractional}. 

Another outcome of our work is the direct connection between bouncing singularities of small black holes in AdS with lightcone singularities on the boundary, $\mathbb{R}\times \mathbb{S}^2$. As $R_h/L \to 0$ the two lines of bouncing singularities in the complex $t/L$ plane move towards the real axis, and coalesce to produce the boundary lightcone singularities at $t = n\pi L$. This, in turn, matches the singularities expected from asymptotic behaviour of pure AdS normal modes. 

Our result also raises a broader question concerning the relation between AdS/CFT and celestial holography. Celestial holography (see \cite{Raclariu:2021zjz, Pasterski:2021rjz} for lecture notes) seeks an operator description of asymptotically flat physics, including an OPE on the celestial sphere. Here, flat black hole dynamics is instead recovered from boundary data organised by the standard AdS/CFT OPE. This suggests that a suitable reorganisation and scaling limit of the AdS boundary OPE could provide dynamical input for the celestial description of black hole backgrounds (see for example \cite{Mitra:2024ugt, deGioia:2024yne, Alday:2024yyj}). 

Our expectation is that this connection could be mediated by the large-order sector of the AdS OPE associated with the first bouncing singularities. The first-return scaling probes the real-time response in an $R_h$-sized neighbourhood of the first complex bouncing singularities, whose distance from the real axis is itself of order $R_h$. The response in this window determines the flat black hole scattering data. Moreover, the expansion near the bouncing singularities is encoded in the large-order behaviour of the AdS short-time coefficients, as illustrated in \cite{Arnaudo:2026axe} and understood through Darboux's principle. This suggests investigating whether resummation of this large-order sector, followed by the flat limit and a transformation to a celestial conformal primary basis, can supply dynamical data for a celestial OPE decomposition in a black hole state. Such a construction would require restoring the angular dependence, going beyond the fixed-$\ell$ procedure we presented here.
Establishing how the boundary operators and their products reorganise in such a limit remains an interesting open direction.

In Section \ref{sec:OPEQNM} we verified that the Schwarzschild-AdS QNMs globally reconstruct the real-time response. Such a decomposition can also be obtained by applying a Prony method for a sufficiently long time window. Restricting to a time window close to the ringdown behaviour, instead, allows for the extraction of the flat QNMs, as shown in Section \ref{sec:extraction}. Starting from the latter localised window and enlarging it produces an intermediate picture, in which fewer flat QNMs can be stably extracted and approximations to the pure AdS normal modes begin to arise. 

This behaviour is reminiscent of the phenomenology of the QNM spectral instability \cite{Nollert:1996rf, Nollert:1998ys, Jaramillo:2020tuu, Cheung:2021bol}.  Indeed, the Schwarzschild-AdS small black hole system studied in this work can be seen as a small $\Lambda$ perturbation of the corresponding flat geometry. This `small' perturbation gives rise to dramatic change in QNM frequencies. However, in the time domain, for short enough times compared to the AdS size the Schwarzschild ringdown is barely affected, and the appropriate short-time description is given by weakly perturbed Schwarzschild QNMs. Only when timescales become large enough to see the scale of the AdS box does the time domain response start to deviate from the Schwarzschild ringdown, recovering the almost-periodic behaviour of the Schwarzschild-AdS solution. This result is similar in spirit to results that show that tidally deformed black holes exhibit small corrections to their QNM frequencies \cite{Giataganas:2026gkl}.

The toy model in Section \ref{sec:toymodel} shows explicitly how resonance poles of an effectively open scattering problem can govern a finite-time response even when the complete system has a real normal-mode spectrum. It would be interesting to investigate this distinction directly in a microscopic CFT. The finite-volume Ising model studied in \cite{Janik:2025zji} provides a concrete example of QNM-like decay followed by exact recurrences. Although its recurrence mechanism differs from the boundary returns studied here, it motivates seeking microscopic models in which an individual returned signal can be reconstructed from exact spectral or OPE data.

\section*{Acknowledgments}
It is a pleasure to thank Javier Carballo and Kostas Skenderis for useful discussions. We would especially like to thank Cristoforo Iossa for discussions surrounding the plane wave basis and insights from the connection procedure with the de Sitter horizon.
PA is supported by the Royal Society grant URF\textbackslash R\textbackslash 231002, `Dynamics of holographic field theories'.
BW is supported by a Royal Society University Research Fellowship and in part by the Science and Technology Facilities Council (Consolidated Grant `New Frontiers in Particle Physics, Cosmology and Gravity'). Parts of the proof in Appendix \ref{app:proof} were developed with assistance from AI, and every step was verified by hand.

\appendix
\addtocontents{toc}{\protect\setcounter{tocdepth}{1}}
\section{Proof of the first-return limit}\label{app:proof}

The relevant ODE in the tortoise coordinate reads
\begin{equation}
\partial_{r_*}^2\phi+\big[\omega^2-V(r)\big]\phi=0.
\end{equation}
We consider the axial Regge--Wheeler problem with $s=\ell=2$. In the proof it is convenient to measure lengths in horizon units:
\begin{equation}
\epsilon=\frac{R_h}{L},\qquad x=\frac{r}{R_h},\qquad \xi=\frac{r_*}{R_h},\qquad \xi_b=\frac{r_*^b}{R_h},\qquad \Omega=\omega R_h.
\end{equation}
We fix the additive constant of the tortoise coordinate by
\begin{equation}
\xi=\xi_\epsilon(x)=\mathrm{PV}\int_0^x\frac{\mathrm{d}y}{1-(1+\epsilon^2)/y+\epsilon^2y^2},
\end{equation}
so that $\xi_\epsilon(x)\to x+\log(x-1)$ on compact subsets of $x>1$. The dimensionless equation is
\begin{equation}
\partial_\xi^2\phi+\big[\Omega^2-V_\epsilon(\xi)\big]\phi=0,
\end{equation}
where
\begin{equation}
V_\epsilon(\xi_\epsilon(x))=R_h^2V(R_hx)=\left(1-\frac{1+\epsilon^2}{x}+\epsilon^2x^2\right)\left(\frac6{x^2}-\frac{3(1+\epsilon^2)}{x^3}\right).
\end{equation}
The AdS boundary is at the finite value $\xi_b$, which tends to infinity as $\epsilon\to0$. To compare with the limiting flat problem, we extend the potential by zero for $\xi>\xi_b$. The solutions there are exactly $e^{\pm\mathrm{i}\Omega\xi}$. This extension is an auxiliary scattering problem: it does not change the ingoing solution or its boundary value and derivative on the physical interval. Its reflection coefficient at finite $\epsilon$ is denoted by $\mathcal R_\ell^{(\epsilon)}$, while $\mathcal R_\ell$ denotes its flat limit. The exterior wavenumber of this auxiliary problem is $\Omega$, rather than $\sqrt{\Omega^2-6\epsilon^2}$; no replacement of one by the other is made at finite $\epsilon$.

\subsection{Convergence of the radial scattering problem}

In what follows, we will only consider $\Omega\in\mathbb{R}$ and write the differential problem as
\begin{equation}
\phi''+\Omega^2\phi=V_{\epsilon}\phi,
\end{equation}
which means that we will be treating $V_{\epsilon}\phi$ as a source to the homogeneous problem in the lhs.
We can then consider the Green kernel $G_0$ for the operator $\partial_\xi^2+\Omega^2$ which vanishes for $v>\xi$ and satisfies $\partial_{\xi}G_0(\xi,v)\bigg|_{\xi=v}=1$, which is 
\begin{equation}
G_0(\xi,v)=\frac{\sin[\Omega(\xi-v)]}{\Omega},\qquad v\le\xi.
\end{equation}
Since $V_{\epsilon}\to 0$ at the horizon $\xi\to-\infty$, the unit-amplitude ingoing solution there satisfies $\phi_{\mathrm{in}}\sim e^{-\mathrm{i}\Omega\xi}$. Applying the Green kernel to the source $V_{\epsilon}\phi_{\mathrm{in}}$ gives the integral equation
\begin{equation}\label{app:ingoingintegral}
\phi_{\mathrm{in}}(\xi)=e^{-\mathrm{i}\Omega\xi}+\int_{-\infty}^{\xi}\frac{\sin[\Omega(\xi-v)]}{\Omega}V_{\epsilon}(v)\phi_{\mathrm{in}}(v)\,\mathrm{d}v.
\end{equation}

Recalling that $V_{\epsilon}(\xi)=0$ for $\xi\ge\xi_b$, we can write the connection formula
\begin{equation}\label{app:connection-extension}
\phi_{\mathrm{in}}(\xi)=C_{\mathrm{in},\mathrm{down}}^{(\epsilon)}e^{-\mathrm{i}\Omega\xi}+C_{\mathrm{in},\mathrm{up}}^{(\epsilon)}e^{\mathrm{i}\Omega\xi},\qquad \xi\geq\xi_b.
\end{equation}
Comparing with \eqref{app:ingoingintegral} and converting the sine function in exponentials, we find
\begin{equation}
C_{\mathrm{in},\mathrm{down}}^{(\epsilon)}=1-\frac{1}{2\mathrm{i}\Omega}\int_{\mathbb R}e^{\mathrm{i}\Omega v}V_{\epsilon}(v)\phi_{\mathrm{in}}(v)\,\mathrm{d}v,\qquad C_{\mathrm{in},\mathrm{up}}^{(\epsilon)}=\frac{1}{2\mathrm{i}\Omega}\int_{\mathbb R}e^{-\mathrm{i}\Omega v}V_{\epsilon}(v)\phi_{\mathrm{in}}(v)\,\mathrm{d}v.
\end{equation}
We define
\begin{equation}
\mathcal R_\ell^{(\epsilon)}(\Omega)=\frac{C_{\mathrm{in},\mathrm{up}}^{(\epsilon)}}{C_{\mathrm{in},\mathrm{down}}^{(\epsilon)}}.
\end{equation}

First, we want to prove that
\begin{equation}
\mathcal R_\ell^{(\epsilon)}(\Omega)\to\mathcal R_\ell(\Omega)\equiv\frac{C_{\mathrm{in},\mathrm{up}}^{\mathrm{flat}}(\Omega)}{C_{\mathrm{in},\mathrm{down}}^{\mathrm{flat}}(\Omega)}
\end{equation}
uniformly on every compact nonzero real-frequency interval. To achieve this, we will prove the convergence of the two coefficients.

Let us show how the proof works for one of the two coefficients, the same will hold for the other one. Let $\mathcal{I}$ be a compact interval not containing 0 and $|\Omega|\ge \delta>0$, $\Omega\in\mathcal{I}$. We have (using H\"older inequality)
\begin{equation}
\begin{aligned}
&\sup_{\Omega\in\mathcal{I}}\left|C_{\mathrm{in},\mathrm{up}}^{(\epsilon)}-C_{\mathrm{in},\mathrm{up}}^{\mathrm{flat}}\right|\le\sup_{\Omega\in\mathcal{I}}\frac{1}{|2\mathrm{i}\Omega|}\int_{\mathbb{R}}\left|e^{-\mathrm{i}\Omega v}\left(V_{\epsilon}\phi_{\mathrm{in}}^{(\epsilon)}(v)-V_{0}\phi_{\mathrm{in}}^{\mathrm{flat}}(v)\right)\right|\mathrm{d}v\le \\
&\frac{1}{2\delta}\sup_{\Omega\in\mathcal{I}}\|e^{-\mathrm{i}\Omega v}\|_{\infty}\|V_{\epsilon}\phi_{\mathrm{in}}^{(\epsilon)}-V_{0}\phi_{\mathrm{in}}^{\mathrm{flat}}\|_1=\frac{1}{2\delta}\sup_{\Omega\in\mathcal{I}}\|V_{\epsilon}\phi_{\mathrm{in}}^{(\epsilon)}-V_{0}\phi_{\mathrm{in}}^{\mathrm{flat}}\|_1.
\end{aligned}
\end{equation}
We then reduce to prove that $\|V_{\epsilon}\phi_{\mathrm{in}}^{(\epsilon)}-V_{0}\phi_{\mathrm{in}}^{\mathrm{flat}}\|_1\to 0$. Let us write
\begin{equation}
V_{\epsilon}\phi_{\mathrm{in}}^{(\epsilon)}-V_{0}\phi_{\mathrm{in}}^{\mathrm{flat}}=\left(V_{\epsilon}-V_{0}\right)\phi_{\mathrm{in}}^{(\epsilon)}+V_{0}\left(\phi_{\mathrm{in}}^{(\epsilon)}-\phi_{\mathrm{in}}^{\mathrm{flat}}\right),
\end{equation}
so that
\begin{equation}
\|V_{\epsilon}\phi_{\mathrm{in}}^{(\epsilon)}-V_{0}\phi_{\mathrm{in}}^{\mathrm{flat}}\|_1\le\|V_{\epsilon}-V_{0}\|_1\|\phi_{\mathrm{in}}^{(\epsilon)}\|_{\infty}+\|V_{0}\|_1\|\phi_{\mathrm{in}}^{(\epsilon)}-\phi_{\mathrm{in}}^{\mathrm{flat}}\|_{\infty}.
\end{equation}
For $0\le\epsilon<1$ the extended potential is nonnegative and
\begin{equation}
\|V_{\epsilon}\|_1=\int_1^\infty\left(\frac6{x^2}-\frac{3(1+\epsilon^2)}{x^3}\right)\mathrm{d}x=\frac{9}{2}-\frac32\epsilon^2\leq\frac{9}{2}.
\end{equation}
Moreover, the regions $1<x<1+\delta$ and $x>X$ contribute at most $6\delta$ and $6/X$, uniformly in $\epsilon$. On every compact interval $1+\delta\leq x\leq X$, the tortoise maps and the potentials converge uniformly to their flat counterparts. It follows that
\begin{equation}
\|V_{\epsilon}-V_0\|_{L^1}\to 0.
\end{equation}

Iterating the integral equation produces ordered integrals $-\infty<v_n<\cdots<v_1<\xi$. Since 
\begin{equation}
\left|\frac{\sin[\Omega(\xi-v)]}{\Omega}\right|\leq\frac1{|\Omega|},
\end{equation}
the $n$th iterate is bounded by
\begin{equation}
\frac1{n!}\left(\frac{\|V_{\epsilon}\|_1}{|\Omega|}\right)^n,
\end{equation}
where the factor $1/n!$ comes from the ordering of the integration variables. Summing the resulting exponential series gives
\begin{equation}
\|\phi_{\mathrm{in}}\|_\infty\leq e^{\|V_{\epsilon}\|_1/|\Omega|}.
\end{equation}

We now compare the iterated series for $V_{\epsilon}$ and $V_0$. The difference of the products of potential factors can be written exactly as
\begin{equation}
\prod_{j=1}^nV_{\epsilon}(v_j)-\prod_{j=1}^nV_0(v_j)=\int_0^1\sum_{j=1}^n[V_{\epsilon}(v_j)-V_0(v_j)]\prod_{k\neq j}[(1-s)V_0(v_k)+sV_{\epsilon}(v_k)]\,\mathrm ds.
\end{equation}
After bounding each Green kernel by $1/\delta$, the resulting sum of absolute-value bounds is symmetric in the integration variables. Its ordered integral is therefore $1/n!$ times the unrestricted integral. Each interpolated potential is nonnegative and has $L^1$ norm at most $9/2$, so we obtain
\begin{equation}
\sup_{\Omega\in\mathcal I}\left\|\phi_{\mathrm{in},n}^{(\epsilon)}-\phi_{\mathrm{in},n}^{\mathrm{flat}}\right\|_\infty
\leq
\frac1{(n-1)!\delta}\left(\frac{9}{2\delta}\right)^{n-1}\|V_{\epsilon}-V_0\|_1.
\end{equation}
Summing over $n\geq1$ gives
\begin{equation}
\sup_{\Omega\in\mathcal I}\left\|\phi_{\mathrm{in}}^{(\epsilon)}-\phi_{\mathrm{in}}^{\mathrm{flat}}\right\|_\infty
\leq
\frac{e^{9/(2\delta)}}{\delta}\|V_{\epsilon}-V_0\|_1\to0.
\end{equation}
Therefore the ingoing solutions converge uniformly for $\Omega\in\mathcal I$.

Hence, the proof of convergence of connection coefficients is done. 

\subsection{Isolating the first return}

With the convention of the main text, the boundary correlator satisfies
\begin{equation}
R_h\widetilde G(\Omega/R_h)=\frac{\phi'_{\mathrm{in}}(\xi_b,\Omega)}{\phi_{\mathrm{in}}(\xi_b,\Omega)},
\end{equation}
where a prime denotes a $\xi$ derivative. Using the connection formula \eqref{app:connection-extension} evaluated at $\xi_b$, we find
\begin{equation}
\phi_{\mathrm{in}}(\xi_b,\Omega)=C_{\mathrm{in},\mathrm{down}}^{(\epsilon)}e^{-\mathrm{i}\Omega\xi_b}\left[1+\mathcal R_\ell^{(\epsilon)}(\Omega)e^{2\mathrm{i}\Omega\xi_b}\right]
\end{equation}
and
\begin{equation}
\phi'_{\mathrm{in}}(\xi_b,\Omega)=-\mathrm{i}\Omega C_{\mathrm{in},\mathrm{down}}^{(\epsilon)}e^{-\mathrm{i}\Omega\xi_b}\left[1-\mathcal R_\ell^{(\epsilon)}(\Omega)e^{2\mathrm{i}\Omega\xi_b}\right].
\end{equation}
Therefore,
\begin{equation}\label{app:exact-boundary-relation}
R_h\widetilde G(\Omega/R_h)=-\mathrm{i}\Omega\frac{1-\mathcal R_\ell^{(\epsilon)}(\Omega)e^{2\mathrm{i}\Omega\xi_b}}{1+\mathcal R_\ell^{(\epsilon)}(\Omega)e^{2\mathrm{i}\Omega\xi_b}}.
\end{equation}

For real $\Omega\neq0$, reality of the potential implies that $\overline{\phi_{\mathrm{in}}}$ is also a solution. The Wronskian
\begin{equation}
W[\overline{\phi_{\mathrm{in}}},\phi_{\mathrm{in}}]=\overline{\phi_{\mathrm{in}}}\phi'_{\mathrm{in}}-\phi_{\mathrm{in}}\overline{\phi'_{\mathrm{in}}}
\end{equation}
is independent of $\xi$. Evaluating it at the horizon and in the free region $\xi\geq\xi_b$ gives
\begin{equation}\label{app:flux-identity}
|C_{\mathrm{in},\mathrm{down}}^{(\epsilon)}|^2-|C_{\mathrm{in},\mathrm{up}}^{(\epsilon)}|^2=1.
\end{equation}
Together with the convergence of the connection coefficients already proved, $|C_{\mathrm{in},\mathrm{down}}^{(\epsilon)}|\geq1$ implies
\begin{equation}
\mathcal R_\ell^{(\epsilon)}(\Omega)\to\mathcal R_\ell(\Omega)=\frac{C_{\mathrm{in},\mathrm{up}}^{\mathrm{flat}}(\Omega)}{C_{\mathrm{in},\mathrm{down}}^{\mathrm{flat}}(\Omega)}
\end{equation}
uniformly on every compact real-frequency interval bounded away from zero. Moreover,
\begin{equation}
1-|\mathcal R_\ell^{(\epsilon)}(\Omega)|^2=\frac1{|C_{\mathrm{in},\mathrm{down}}^{(\epsilon)}|^2}>0.
\end{equation}
Hence $|\mathcal R_\ell^{(\epsilon)}(\Omega)|<1$, the boundary denominator cannot vanish, and the geometric expansion is convergent:
\begin{equation}\label{app:return-expansion}
R_h\widetilde G(\Omega/R_h)=-\mathrm{i}\Omega+2\mathrm{i}\Omega\sum_{m\geq1}(-1)^{m-1}\left[\mathcal R_\ell^{(\epsilon)}(\Omega)\right]^m e^{2\mathrm{i}m\Omega\xi_b}.
\end{equation}
Evaluating the same Wronskian directly at $\xi_b$ also gives
\begin{equation}\label{app:positive-weight}
-\frac{R_h\,\mathrm{Im}\,\widetilde G(\Omega/R_h)}{\Omega}=\frac1{|\phi_{\mathrm{in}}(\xi_b,\Omega)|^2}>0.
\end{equation}

Since $R_h\xi_b=r_*^b$, the phase of the $m$th term corresponds to the physical delay $2mr_*^b$. Near the first return we set $t=2r_*^b+R_hT$. Multiplying \eqref{app:return-expansion} by the corresponding Fourier phase gives
\begin{equation}
\begin{aligned}
&e^{-\mathrm{i}\Omega(2\xi_b+T)}R_h\widetilde G(\Omega/R_h)\\
&\quad=e^{-\mathrm{i}\Omega T}\left[-\mathrm{i}\Omega e^{-2\mathrm{i}\Omega\xi_b}+2\mathrm{i}\Omega\mathcal R_\ell^{(\epsilon)}(\Omega)+2\mathrm{i}\Omega\sum_{m\geq2}(-1)^{m-1}\left[\mathcal R_\ell^{(\epsilon)}(\Omega)\right]^m e^{2\mathrm{i}(m-1)\Omega\xi_b}\right].
\end{aligned}
\end{equation}
The $m=1$ term is the unique return term for which the phase proportional to $\xi_b$ cancels. The prompt term gives $\delta'(t)$, while all later returns retain a rapidly oscillating phase. To justify the limit after Fourier integration, we must control both small and large real frequencies.

\subsection{A real-frequency representation of the returned waveform}

Changing variables from $\omega$ to $\Omega=\omega R_h$ in the retarded Fourier transform gives
\begin{equation}
R_h^2G(t)=\int_{\mathbb R}\frac{\mathrm d\Omega}{2\pi}e^{-\mathrm{i}\Omega t/R_h}R_h\widetilde G(\Omega/R_h).
\end{equation}
We will obtain an absolutely convergent representation for $t>0$ and then take the first-return limit directly in this expression.

\subsubsection{Around $\Omega=0$}

Here $\epsilon>0$ is fixed, so $\xi_b$ is finite. Near the horizon $x=1$,
\begin{equation}
\xi=\frac{\log(x-1)}{1+3\epsilon^2}+\mathcal O(1)\qquad V_{\epsilon}=\mathcal O(x-1).
\end{equation}
as $x\to1^+$. 
Therefore
\begin{equation}
V_{\epsilon}(\xi)=\mathcal O\left(e^{(1+3\epsilon^2)\xi}\right),\qquad \xi\to-\infty.
\end{equation}
Consequently,
\begin{equation}
\int_{-\infty}^{\xi_b}(1+\xi_b-v)V_\epsilon(v)\,\mathrm dv<\infty.
\end{equation}
The integral kernel has the continuous limit
\begin{equation}
\frac{\sin[\Omega(\xi-v)]}{\Omega}\to\xi-v,\qquad
\left|\frac{\sin[\Omega(\xi-v)]}{\Omega}\right|\leq\xi-v,\qquad v\leq\xi.
\end{equation}
At any $\xi\leq\xi_b$, its $n$th ordered iterate is bounded by
\begin{equation}
\frac1{n!}\left[\int_{-\infty}^{\xi_b}(\xi_b-v)V_\epsilon(v)\,\mathrm dv\right]^n.
\end{equation}
These summable bounds allow the limit $\Omega\to0$ to be taken term by term and show that $\phi_{\mathrm{in}}(\xi_b,\Omega)$ is continuous at zero. Differentiating the integral equation in $\xi$ gives
\begin{equation}
\phi'_{\mathrm{in}}(\xi,\Omega)=-\mathrm{i}\Omega e^{-\mathrm{i}\Omega\xi}+\int_{-\infty}^{\xi}\cos[\Omega(\xi-v)]V_\epsilon(v)\phi_{\mathrm{in}}(v,\Omega)\,\mathrm dv.
\end{equation}
The same bounds make the integrand uniformly dominated by an integrable function for $\xi=\xi_b$, so $\phi'_{\mathrm{in}}(\xi_b,\Omega)$ is also continuous at zero.

Setting $\Omega=0$ in these equations gives
\begin{equation}
\phi_{\mathrm{in}}(\xi,0)=1+\int_{-\infty}^{\xi}(\xi-v)V_\epsilon(v)\phi_{\mathrm{in}}(v,0)\,\mathrm dv\geq1,
\end{equation}
\begin{equation}
\phi'_{\mathrm{in}}(\xi,0)=\int_{-\infty}^{\xi}V_\epsilon(v)\phi_{\mathrm{in}}(v,0)\,\mathrm dv\geq0.
\end{equation}
The inequalities follow by iterating from the positive free solution $1$. Thus, the correlator is finite at zero frequency, with
\begin{equation}\label{app:static-value}
R_h\widetilde G(0)=\frac{\phi'_{\mathrm{in}}(\xi_b,0)}{\phi_{\mathrm{in}}(\xi_b,0)}\geq0.
\end{equation}
In addition, \eqref{app:positive-weight} shows that $-R_h\,\mathrm{Im}\,\widetilde G(\Omega/R_h)/\Omega$ has a finite continuous limit at zero, so $R_h\,\mathrm{Im}\,\widetilde G(\Omega/R_h)$ vanishes linearly with $\Omega$.

\subsubsection{Large frequency behaviour}

We next determine the behaviour at large real frequency, with estimates uniform for $0<\epsilon\leq1$. As a function of $x$,
\begin{equation}
V_\epsilon=6\epsilon^2-\frac{3\epsilon^2(1+\epsilon^2)}x+\frac6{x^2}-\frac{9(1+\epsilon^2)}{x^3}+\frac{3(1+\epsilon^2)^2}{x^4},
\end{equation}
so
\begin{equation}
\partial_xV_\epsilon=\frac{3\epsilon^2(1+\epsilon^2)}{x^2}-\frac{12}{x^3}+\frac{27(1+\epsilon^2)}{x^4}-\frac{12(1+\epsilon^2)^2}{x^5}.
\end{equation}
Integrating the absolute values of the individual terms gives
\begin{equation}
\int_{-\infty}^{\xi_b}|\partial_\xi V_\epsilon|\,\mathrm d\xi=\int_1^\infty|\partial_xV_\epsilon|\,\mathrm dx\leq18+18\epsilon^2+6\epsilon^4\leq42.
\end{equation}
The endpoint values are
\begin{equation}
V_\epsilon(-\infty)=0,\qquad V_\epsilon(\xi_b^-)=6\epsilon^2.
\end{equation}
We will also use
\begin{equation}
\|V_\epsilon\|_\infty+\|\partial_\xi V_\epsilon\|_\infty+\|\partial_\xi^2V_\epsilon\|_1\leq C,
\end{equation}
where these norms and derivatives are taken on the physical interval $(-\infty,\xi_b)$, before the zero extension, and $C$ is independent of $\epsilon$. To check this, observe that
\begin{equation}
\partial_\xi V_\epsilon=\left(1-\frac{1+\epsilon^2}{x}+\epsilon^2x^2\right)\partial_xV_\epsilon
\end{equation}
is a polynomial in $1/x$ with uniformly bounded coefficients, including a constant term. Its $x$ derivative is a sum of inverse powers $x^{-j}$ with $j\geq2$, and therefore
\begin{equation}
\int_{-\infty}^{\xi_b}|\partial_\xi^2V_\epsilon|\,\mathrm d\xi=\int_1^\infty|\partial_x(\partial_\xi V_\epsilon)|\,\mathrm dx\leq C.
\end{equation}
Also, $\partial_\xi V_\epsilon\to0$ at the horizon.

For real $\Omega\neq0$, multiplying the integral equation by $e^{\mathrm{i}\Omega\xi}$ gives
\begin{equation}\label{app:weighted-integral}
e^{\mathrm{i}\Omega\xi}\phi_{\mathrm{in}}(\xi)=1+\frac1{2\mathrm{i}\Omega}\int_{-\infty}^{\xi}\left[e^{2\mathrm{i}\Omega(\xi-v)}-1\right]V_\epsilon(v)e^{\mathrm{i}\Omega v}\phi_{\mathrm{in}}(v)\,\mathrm dv.
\end{equation}
Since the kernel has modulus at most $1/|\Omega|$, the same ordered-iteration estimate as before gives
\begin{equation}
\left\|e^{\mathrm{i}\Omega\xi}\phi_{\mathrm{in}}\right\|_\infty\leq e^{9/(2|\Omega|)},\qquad
\left\|e^{\mathrm{i}\Omega\xi}\phi_{\mathrm{in}}-1\right\|_\infty\leq\frac9{2|\Omega|}e^{9/(2|\Omega|)}=\mathcal O(|\Omega|^{-1}),
\end{equation}
uniformly in $\epsilon$.

We now differentiate this integral equation and integrate by parts:
\begin{equation}
\begin{aligned}
\partial_\xi\left(e^{\mathrm{i}\Omega\xi}\phi_{\mathrm{in}}(\xi)\right)&=\int_{-\infty}^{\xi}e^{2\mathrm{i}\Omega(\xi-v)}V_\epsilon(v)e^{\mathrm{i}\Omega v}\phi_{\mathrm{in}}(v)\,\mathrm dv\\
&=-\frac{V_\epsilon(\xi)e^{\mathrm{i}\Omega\xi}\phi_{\mathrm{in}}(\xi)}{2\mathrm{i}\Omega}+\frac1{2\mathrm{i}\Omega}\int_{-\infty}^{\xi}e^{2\mathrm{i}\Omega(\xi-v)}\partial_v\left[V_\epsilon(v)e^{\mathrm{i}\Omega v}\phi_{\mathrm{in}}(v)\right]\,\mathrm dv.
\end{aligned}
\end{equation}
The first expression makes the derivative bounded. Splitting the derivative inside the second integral and using the bounds above gives, for $|\Omega|\geq1$,
\begin{equation}
\left\|\partial_\xi(e^{\mathrm{i}\Omega\xi}\phi_{\mathrm{in}})\right\|_\infty
\leq\frac{C}{|\Omega|}+\frac9{4|\Omega|}\left\|\partial_\xi(e^{\mathrm{i}\Omega\xi}\phi_{\mathrm{in}})\right\|_\infty.
\end{equation}
For sufficiently large $|\Omega|$, the last term can be moved to the left-hand side. Hence $\|\partial_\xi(e^{\mathrm{i}\Omega\xi}\phi_{\mathrm{in}})\|_\infty=\mathcal O(|\Omega|^{-1})$ uniformly in $\epsilon$.

We can now estimate the remaining integral more sharply. Its term containing $V_\epsilon\,\partial_v(e^{\mathrm{i}\Omega v}\phi_{\mathrm{in}})$ is $\mathcal O(|\Omega|^{-1})$ by the last bound and $\|V_\epsilon\|_1\leq9/2$. For the term containing $\partial_vV_\epsilon$, a further integration by parts gives
\begin{equation}
\begin{aligned}
&\left|\int_{-\infty}^{\xi_b}e^{2\mathrm{i}\Omega(\xi_b-v)}\partial_vV_\epsilon(v)e^{\mathrm{i}\Omega v}\phi_{\mathrm{in}}(v)\,\mathrm dv\right|\\
&\quad\leq\frac1{2|\Omega|}\left[\left(\|\partial_\xi V_\epsilon\|_\infty+\|\partial_\xi^2V_\epsilon\|_1\right)\left\|e^{\mathrm{i}\Omega\xi}\phi_{\mathrm{in}}\right\|_\infty
+\|\partial_\xi V_\epsilon\|_1\left\|\partial_\xi(e^{\mathrm{i}\Omega\xi}\phi_{\mathrm{in}})\right\|_\infty\right]
=\mathcal O(|\Omega|^{-1}).
\end{aligned}
\end{equation}
The horizon boundary term vanishes; the other boundary term is included in $\|\partial_\xi V_\epsilon\|_\infty$. Thus, at $\xi_b$, the integral remainder in the preceding formula for $\partial_\xi(e^{\mathrm{i}\Omega\xi}\phi_{\mathrm{in}})$ is $\mathcal O(|\Omega|^{-2})$, including its prefactor $1/(2\mathrm{i}\Omega)$. Since $e^{\mathrm{i}\Omega\xi_b}\phi_{\mathrm{in}}(\xi_b)=1+\mathcal O(|\Omega|^{-1})$ is bounded away from zero, division by this factor gives
\begin{equation}\label{app:uv-refined}
R_h\widetilde G(\Omega/R_h)=\frac{\phi'_{\mathrm{in}}(\xi_b,\Omega)}{\phi_{\mathrm{in}}(\xi_b,\Omega)}
=-\mathrm{i}\Omega-\frac{6\epsilon^2}{2\mathrm{i}\Omega}+\mathcal O(|\Omega|^{-2}).
\end{equation}
uniformly as $|\Omega|\to\infty$ on the real axis. Thus the expression
\begin{equation}\label{app:integrable-subtraction}
R_h\widetilde G(\Omega/R_h)+\mathrm{i}\Omega-\frac{3\epsilon^2}{1-\mathrm{i}\Omega}
\end{equation}
is $\mathcal O(|\Omega|^{-2})$. Taking the imaginary part gives
\begin{equation}\label{app:uv-integrable}
R_h\,\mathrm{Im}\,\widetilde G(\Omega/R_h)+\Omega-\frac{3\epsilon^2\Omega}{1+\Omega^2}=\mathcal O(|\Omega|^{-2}).
\end{equation}
Finally, the exact boundary relation \eqref{app:exact-boundary-relation} implies
\begin{equation}
|\mathcal R_\ell^{(\epsilon)}(\Omega)|=\left|\frac{\mathrm{i}\Omega+R_h\widetilde G(\Omega/R_h)}{\mathrm{i}\Omega-R_h\widetilde G(\Omega/R_h)}\right|\leq\frac{C\epsilon^2}{\Omega^2}+\frac{C}{|\Omega|^3}.
\end{equation}
Taking $\epsilon\to0$ at fixed nonzero real $\Omega$, and using the convergence of the scattering coefficients already proved, therefore gives
\begin{equation}\label{app:flat-refined-uv}
|\mathcal R_\ell(\Omega)|\leq\frac{C}{|\Omega|^3}
\end{equation}
for sufficiently large $|\Omega|$.

\subsubsection{Direct inversion}

For fixed $\epsilon>0$, the expression \eqref{app:integrable-subtraction} is continuous on the real axis and belongs to $L^1(\mathbb R)$. Its inverse Fourier transform is therefore continuous. It is also causal: $\widetilde G$ is the retarded boundary response, while the two subtracted terms correspond to the prompt contact term and a causal exponential. Explicitly, for $t>0$,
\begin{equation}
R_h^2G(t)=3\epsilon^2e^{-t/R_h}+\int_{\mathbb R}\frac{\mathrm d\Omega}{2\pi}e^{-\mathrm{i}\Omega t/R_h}\left[R_h\widetilde G(\Omega/R_h)+\mathrm{i}\Omega-\frac{3\epsilon^2}{1-\mathrm{i}\Omega}\right].
\end{equation}
Reality of the time-domain response gives $\widetilde G(-\omega)=\overline{\widetilde G(\omega)}$ on the real axis. The complex conjugate of the square bracket consequently has the time-reversed inverse transform, which vanishes for $t>0$. Subtracting it from the square bracket gives the exact, absolutely convergent representation
\begin{equation}\label{app:real-frequency-inversion}
R_h^2G(t)=3\epsilon^2e^{-t/R_h}+\frac{\mathrm{i}}{\pi}\int_{\mathbb R}e^{-\mathrm{i}\Omega t/R_h}\left[R_h\,\mathrm{Im}\,\widetilde G(\Omega/R_h)+\Omega-\frac{3\epsilon^2\Omega}{1+\Omega^2}\right]\,\mathrm d\Omega,\qquad t>0.
\end{equation}
The prompt $\delta'(t)$ is omitted because $t>0$.

\subsection{Taking the first-return limit}

We now set
\begin{equation}
t=2r_*^b+R_hT,\qquad \frac{t}{R_h}=2\xi_b+T.
\end{equation}
Since $\xi_b\to\infty$ as $\epsilon\to0$, this time is positive for sufficiently small $\epsilon$, uniformly for $T$ in any fixed compact interval. We can therefore use \eqref{app:real-frequency-inversion}. Fix $0<\delta<K$ and split the frequency axis into
\begin{equation}
|\Omega|<\delta,\qquad \delta\leq|\Omega|\leq K,\qquad |\Omega|>K.
\end{equation}
The proposed limiting expression is
\begin{equation}
\frac{\mathrm{i}}{\pi}\int_{\mathbb R}\Omega e^{-\mathrm{i}\Omega T}\mathcal R_\ell(\Omega)\,\mathrm d\Omega.
\end{equation}
It is absolutely convergent, since $|\mathcal R_\ell|<1$ on the nonzero real axis and \eqref{app:flat-refined-uv} controls infinity. We compare the two integrals in the same three regions. The explicit term $3\epsilon^2e^{-(2\xi_b+T)}$ in \eqref{app:real-frequency-inversion} tends uniformly to zero.

\subsubsection{The limit in the middle region}

The flat problem satisfies the same identity \eqref{app:flux-identity}, so $|\mathcal R_\ell(\Omega)|<1$ for real $\Omega\neq0$. Continuity and uniform convergence on $\delta\leq|\Omega|\leq K$ imply that, for sufficiently small $\epsilon$, there is a constant $q<1$ such that
\begin{equation}
|\mathcal R_\ell^{(\epsilon)}(\Omega)|\leq q,\qquad \delta\leq|\Omega|\leq K.
\end{equation}
Taking the imaginary part of \eqref{app:return-expansion} gives
\begin{equation}\label{app:imaginary-return-expansion}
R_h\,\mathrm{Im}\,\widetilde G(\Omega/R_h)+\Omega=\Omega\sum_{m\geq1}(-1)^{m-1}\left[\left(\mathcal R_\ell^{(\epsilon)}(\Omega)\right)^m e^{2\mathrm{i}m\Omega\xi_b}+\left(\overline{\mathcal R_\ell^{(\epsilon)}(\Omega)}\right)^m e^{-2\mathrm{i}m\Omega\xi_b}\right].
\end{equation}
The remainder after the first $N$ terms is bounded uniformly by
\begin{equation}
\frac{2Kq^{N+1}}{1-q}.
\end{equation}
Its integral on the finite middle region can therefore be made arbitrarily small by choosing $N$ large enough.

For fixed $N$, uniform convergence allows us to replace $\mathcal R_\ell^{(\epsilon)}$ by $\mathcal R_\ell$ in each term, with an error tending to zero. After multiplication by the first-return phase, the $m=1$ term in the first part of \eqref{app:imaginary-return-expansion} becomes
\begin{equation}
e^{-\mathrm{i}\Omega(2\xi_b+T)}\Omega\mathcal R_\ell(\Omega)e^{2\mathrm{i}\Omega\xi_b}=\Omega\mathcal R_\ell(\Omega)e^{-\mathrm{i}\Omega T}.
\end{equation}
The remaining phases are
\begin{equation}
e^{\mathrm{i}\Omega[2(m-1)\xi_b-T]},\qquad m\geq2,
\end{equation}
in the first part, and
\begin{equation}
e^{-\mathrm{i}\Omega[2(m+1)\xi_b+T]},\qquad m\geq1,
\end{equation}
in the conjugate part. Their oscillation parameters tend uniformly to infinity for bounded $T$.

To see directly why these integrals vanish, approximate each continuous coefficient on the middle region in $L^1$ by a step function. The error in the oscillatory integral is bounded by this $L^1$ error. On any interval $[a,b]$ of the step function,
\begin{equation}
\left|\int_a^b e^{\mathrm{i}A\Omega}\,\mathrm d\Omega\right|=\left|\frac{e^{\mathrm{i}Ab}-e^{\mathrm{i}Aa}}{\mathrm{i}A}\right|\leq\frac2{|A|}.
\end{equation}
Thus each of the remaining terms vanishes uniformly on compact intervals of $T$. The rational term in \eqref{app:real-frequency-inversion} contributes at most $3\epsilon^2\pi^{-1}\int_{\delta\leq|\Omega|\leq K}|\Omega|/(1+\Omega^2)\,\mathrm d\Omega$, which also tends to zero. Taking first $\epsilon\to0$ at fixed $N$ and then $N\to\infty$ gives
\begin{equation}\label{app:middle-limit}
\begin{aligned}
&\frac{\mathrm{i}}{\pi}\int_{\delta\leq|\Omega|\leq K}e^{-\mathrm{i}\Omega(2\xi_b+T)}\left[R_h\,\mathrm{Im}\,\widetilde G(\Omega/R_h)+\Omega-\frac{3\epsilon^2\Omega}{1+\Omega^2}\right]\,\mathrm d\Omega\\
&\qquad\to\frac{\mathrm{i}}{\pi}\int_{\delta\leq|\Omega|\leq K}\Omega e^{-\mathrm{i}\Omega T}\mathcal R_\ell(\Omega)\,\mathrm d\Omega.
\end{aligned}
\end{equation}

\subsubsection{The limit in the high-frequency regions}

The uniform estimate \eqref{app:uv-integrable} gives, for sufficiently large $K$,
\begin{equation}
\left|\frac{\mathrm{i}}{\pi}\int_{|\Omega|>K}e^{-\mathrm{i}\Omega(2\xi_b+T)}\left[R_h\,\mathrm{Im}\,\widetilde G(\Omega/R_h)+\Omega-\frac{3\epsilon^2\Omega}{1+\Omega^2}\right]\,\mathrm d\Omega\right|\leq\frac{2C_1}{\pi K},
\end{equation}
with $C_1$ independent of sufficiently small $\epsilon$. Similarly, \eqref{app:flat-refined-uv} gives
\begin{equation}
\left|\frac{\mathrm{i}}{\pi}\int_{|\Omega|>K}\Omega e^{-\mathrm{i}\Omega T}\mathcal R_\ell(\Omega)\,\mathrm d\Omega\right|\leq\frac{2C_2}{\pi K}.
\end{equation}
Both tails are therefore uniformly small when $K$ is large.

\subsubsection{The limit in the low-frequency region}

For the small-frequency estimate we prove and use the following identity 
\begin{equation}\label{app:static-sum-rule}
\pi R_h\widetilde G(0)+\int_{\mathbb R}\left[-\frac{R_h\,\mathrm{Im}\,\widetilde G(\Omega/R_h)}{\Omega}-1\right]\,\mathrm d\Omega=0.
\end{equation}
Here the quotient is continuous at zero by \eqref{app:positive-weight}, and the integrand is integrable at infinity by \eqref{app:uv-refined}. Since \eqref{app:integrable-subtraction} belongs to $L^1(\mathbb R)$ and its inverse transform is causal, for every $s>0$,
\begin{equation}
\int_{\mathbb R}e^{\mathrm{i}\Omega s}\left[R_h\widetilde G(\Omega/R_h)+\mathrm{i}\Omega-\frac{3\epsilon^2}{1-\mathrm{i}\Omega}\right]\,\mathrm d\Omega=0.
\end{equation}
Moreover, for every $\eta>0$,
\begin{equation}
\int_0^\infty\int_{\mathbb R}e^{-\eta s}\left|R_h\widetilde G(\Omega/R_h)+\mathrm{i}\Omega-\frac{3\epsilon^2}{1-\mathrm{i}\Omega}\right|\,\mathrm d\Omega\,\mathrm ds
=\frac1\eta\int_{\mathbb R}\left|R_h\widetilde G(\Omega/R_h)+\mathrm{i}\Omega-\frac{3\epsilon^2}{1-\mathrm{i}\Omega}\right|\,\mathrm d\Omega<\infty.
\end{equation}
Using
\begin{equation}
\frac1{\eta-\mathrm{i}\Omega}=\int_0^\infty e^{-\eta s}e^{\mathrm{i}\Omega s}\,\mathrm ds,\qquad \eta>0,
\end{equation}
and exchanging the order of integration using the absolute integrability just shown, we find 
\begin{equation}
\begin{aligned}
&\int_{\mathbb R}\frac{R_h\widetilde G(\Omega/R_h)+\mathrm{i}\Omega-3\epsilon^2/(1-\mathrm{i}\Omega)}{\eta-\mathrm{i}\Omega}\,\mathrm d\Omega\\
&=\int_0^\infty e^{-\eta s}\left\{\int_{\mathbb R}e^{\mathrm{i}\Omega s}\left[R_h\widetilde G(\Omega/R_h)+\mathrm{i}\Omega-\frac{3\epsilon^2}{1-\mathrm{i}\Omega}\right]\,\mathrm d\Omega\right\}\,\mathrm ds=0.
\end{aligned}
\end{equation}

Using $(\eta-\mathrm{i}\Omega)^{-1}=(\eta+\mathrm{i}\Omega)/(\eta^2+\Omega^2)$ and taking the real part gives
\begin{equation}
\begin{aligned}
&\int_{\mathbb R}\frac{\eta}{\eta^2+\Omega^2}\left[R_h\,\mathrm{Re}\,\widetilde G(\Omega/R_h)-\frac{3\epsilon^2}{1+\Omega^2}\right]\,\mathrm d\Omega=\int_{\mathbb R}\frac{\Omega^2}{\eta^2+\Omega^2}\left[\frac{R_h\,\mathrm{Im}\,\widetilde G(\Omega/R_h)}{\Omega}+1-\frac{3\epsilon^2}{1+\Omega^2}\right]\,\mathrm d\Omega.
\end{aligned}
\end{equation}
We now take $\eta\to0^+$ on both sides.

On the left, the substitution $\Omega=\eta v$ gives
\begin{equation}
\begin{aligned}
&\int_{\mathbb R}\frac{\eta}{\eta^2+\Omega^2}\left[R_h\,\mathrm{Re}\,\widetilde G(\Omega/R_h)-\frac{3\epsilon^2}{1+\Omega^2}\right]\,\mathrm d\Omega=\int_{\mathbb R}\frac{R_h\,\mathrm{Re}\,\widetilde G(\eta v/R_h)-3\epsilon^2/(1+\eta^2v^2)}{1+v^2}\,\mathrm dv\\
&\qquad\to [R_h\widetilde G(0)-3\epsilon^2]\int_{\mathbb R}\frac{\mathrm dv}{1+v^2}
=\pi[R_h\widetilde G(0)-3\epsilon^2],
\end{aligned}
\end{equation}
where $\widetilde G(0)$ is real. At fixed $\epsilon$, continuity on the real axis and \eqref{app:uv-refined} imply that $R_h\,\mathrm{Re}\,\widetilde G(\Omega/R_h)$ is bounded. Hence
\begin{equation}
\left|\frac{R_h\,\mathrm{Re}\,\widetilde G(\eta v/R_h)-3\epsilon^2/(1+\eta^2v^2)}{1+v^2}\right|\leq\frac{C}{1+v^2},
\end{equation}
with $C$ independent of $\eta$ and $v$. This integrable bound justifies the limit by dominated convergence.

On the right, from the identity \eqref{app:positive-weight} it follows
\begin{equation}
\frac{R_h\,\mathrm{Im}\,\widetilde G(\Omega/R_h)}{\Omega}
=-\frac1{|\phi_{\mathrm{in}}(\xi_b,\Omega)|^2},
\end{equation}
so the square bracket is continuous at zero. At infinity, \eqref{app:uv-integrable} shows that it is $\mathcal O(|\Omega|^{-3})$. Consequently, for a constant $C$ depending on the fixed $\epsilon$ but not on $\eta$,
\begin{equation}
\left|\frac{\Omega^2}{\eta^2+\Omega^2}\left[\frac{R_h\,\mathrm{Im}\,\widetilde G(\Omega/R_h)}{\Omega}+1-\frac{3\epsilon^2}{1+\Omega^2}\right]\right|
\leq\left|\frac{R_h\,\mathrm{Im}\,\widetilde G(\Omega/R_h)}{\Omega}+1-\frac{3\epsilon^2}{1+\Omega^2}\right|
\leq\frac{C}{1+\Omega^2}.
\end{equation}
Dominated convergence therefore yields
\begin{equation}
\pi[R_h\widetilde G(0)-3\epsilon^2]
=\int_{\mathbb R}\left[\frac{R_h\,\mathrm{Im}\,\widetilde G(\Omega/R_h)}{\Omega}+1-\frac{3\epsilon^2}{1+\Omega^2}\right]\,\mathrm d\Omega.
\end{equation}
Finally, 
\begin{equation}
\int_{\mathbb R}3\epsilon^2(1+\Omega^2)^{-1}\,\mathrm d\Omega=3\pi\epsilon^2,
\end{equation} 
so these terms cancel between the two sides and give
\begin{equation}
\pi R_h\widetilde G(0)+\int_{\mathbb R}\left[-\frac{R_h\,\mathrm{Im}\,\widetilde G(\Omega/R_h)}{\Omega}-1\right]\,\mathrm d\Omega=0,
\end{equation}
which proves \eqref{app:static-sum-rule}.

We next show
\begin{equation}\label{app:outside-zero}
\int_{|\Omega|\geq\delta}\left[-\frac{R_h\,\mathrm{Im}\,\widetilde G(\Omega/R_h)}{\Omega}-1\right]\,\mathrm d\Omega\to0.
\end{equation}
On $\delta\leq|\Omega|\leq K$, dividing \eqref{app:imaginary-return-expansion} by $-\Omega$ gives
\begin{equation}
-\frac{R_h\,\mathrm{Im}\,\widetilde G(\Omega/R_h)}{\Omega}-1=-\sum_{m\geq1}(-1)^{m-1}\left[\left(\mathcal R_\ell^{(\epsilon)}(\Omega)\right)^m e^{2\mathrm{i}m\Omega\xi_b}+\left(\overline{\mathcal R_\ell^{(\epsilon)}(\Omega)}\right)^m e^{-2\mathrm{i}m\Omega\xi_b}\right].
\end{equation}
All terms then retain one of the phases $e^{\pm2\mathrm{i}m\Omega\xi_b}$, with $m\geq1$. The same truncation and step-function argument as above makes their integral tend to zero. On $|\Omega|>K$, \eqref{app:uv-refined} bounds the integrand by $C/\Omega^2$, uniformly in $\epsilon$. Taking first $\epsilon\to0$ and then $K\to\infty$ proves \eqref{app:outside-zero}.

Splitting the static identity \eqref{app:static-sum-rule} at $|\Omega|=\delta$ now gives
\begin{equation}
\pi R_h\widetilde G(0)+\int_{-\delta}^{\delta}\left[-\frac{R_h\,\mathrm{Im}\,\widetilde G(\Omega/R_h)}{\Omega}\right]\,\mathrm d\Omega=2\delta-\int_{|\Omega|\geq\delta}\left[-\frac{R_h\,\mathrm{Im}\,\widetilde G(\Omega/R_h)}{\Omega}-1\right]\,\mathrm d\Omega\to2\delta.
\end{equation}
Both terms on the left are nonnegative, by \eqref{app:static-value} and \eqref{app:positive-weight}. In particular,
\begin{equation}
\limsup_{\epsilon\to0}\int_{-\delta}^{\delta}\left[-\frac{R_h\,\mathrm{Im}\,\widetilde G(\Omega/R_h)}{\Omega}\right]\,\mathrm d\Omega\leq2\delta.
\end{equation}
Using positivity and $|e^{-\mathrm{i}\Omega t/R_h}|=1$, the low-frequency part of the direct inversion satisfies
\begin{equation}
\begin{aligned}
&\left|\frac{\mathrm{i}}{\pi}\int_{-\delta}^{\delta}e^{-\mathrm{i}\Omega t/R_h}\left[R_h\,\mathrm{Im}\,\widetilde G(\Omega/R_h)+\Omega-\frac{3\epsilon^2\Omega}{1+\Omega^2}\right]\,\mathrm d\Omega\right|\\
&\qquad\leq\frac{\delta}{\pi}\int_{-\delta}^{\delta}\left[-\frac{R_h\,\mathrm{Im}\,\widetilde G(\Omega/R_h)}{\Omega}\right]\,\mathrm d\Omega+\frac{\delta^2}{\pi}+\frac{3\epsilon^2}{\pi}\log(1+\delta^2).
\end{aligned}
\end{equation}
Here $\int_{-\delta}^{\delta}|\Omega|\,\mathrm d\Omega=\delta^2$ and $\int_{-\delta}^{\delta}|\Omega|/(1+\Omega^2)\,\mathrm d\Omega=\log(1+\delta^2)$. Consequently,
\begin{equation}\label{app:finite-low-bound}
\limsup_{\epsilon\to0}\sup_{t>0}\left|\frac{\mathrm{i}}{\pi}\int_{-\delta}^{\delta}e^{-\mathrm{i}\Omega t/R_h}\left[R_h\,\mathrm{Im}\,\widetilde G(\Omega/R_h)+\Omega-\frac{3\epsilon^2\Omega}{1+\Omega^2}\right]\,\mathrm d\Omega\right|\leq\frac{3\delta^2}{\pi}.
\end{equation}
The flat integral is controlled directly by $|\mathcal R_\ell|<1$:
\begin{equation}
\left|\frac{\mathrm{i}}{\pi}\int_{|\Omega|<\delta}\Omega e^{-\mathrm{i}\Omega T}\mathcal R_\ell(\Omega)\,\mathrm d\Omega\right|\leq\frac{\delta^2}{\pi}.
\end{equation}
Thus both low-frequency contributions vanish uniformly as $\delta\to0$.

\subsubsection{Conclusion of the proof}

For fixed $0<\delta<K$, the middle-frequency difference tends uniformly to zero on compact intervals of $T$. The two low-frequency contributions are bounded in the limit by $3\delta^2/\pi$ and $\delta^2/\pi$, while the high-frequency tails are bounded by $2(C_1+C_2)/(\pi K)$. The explicit term $3\epsilon^2e^{-(2\xi_b+T)}$ also vanishes. Hence
\begin{equation}
\limsup_{\epsilon\to0}\left|R_h^2G(2r_*^b+R_hT)-\frac{\mathrm{i}}{\pi}\int_{\mathbb R}\Omega e^{-\mathrm{i}\Omega T}\mathcal R_\ell(\Omega)\,\mathrm d\Omega\right|\leq\frac{4\delta^2}{\pi}+\frac{C}{K},
\end{equation}
uniformly on every compact real interval of $T$. Sending $\delta\to0$ and $K\to\infty$ proves
\begin{equation}\label{app:main}
\lim_{R_h/L\to0^+}R_h^2G(2r_*^b+R_hT)=\int_{\mathbb R}\frac{\mathrm d\Omega}{2\pi}\,2\mathrm{i}\Omega\,\mathcal R_\ell(\Omega)e^{-\mathrm{i}\Omega T},\qquad \mathcal R_\ell(\Omega)=\frac{C_{\mathrm{in},\mathrm{up}}^{\mathrm{flat}}(\Omega)}{C_{\mathrm{in},\mathrm{down}}^{\mathrm{flat}}(\Omega)}.
\end{equation}
The convergence is uniform for $T$ in every compact real interval.

\bibliographystyle{ytphys}
\bibliography{refs}

\end{document}